\documentclass[bibyear]{aa} 

\usepackage{lscape}
\usepackage{float}
\usepackage{graphicx}
\usepackage{txfonts}
\begin{document} 

\title{The Seahorse Nebula: New views of the filamentary infrared dark cloud G304.74+01.32 from SABOCA, \textit{Herschel}, and \textit{WISE}\thanks{This publication is based on data 
acquired with the Atacama Pathfinder EXperiment (APEX) under programmes 
{\tt 083.F-9302(A)} and {\tt 089.F-9310(A)}. APEX is a collaboration between the 
Max-Planck-Institut f\"{u}r Radioastronomie, the European Southern 
Observatory, and the Onsala Space Observatory.}}

   \author{O.~Miettinen\inst{1,2,3}}

   \institute{Avarea Oy, Rautatiel\"{a}isenkatu 6, FI-00520 Helsinki, Finland \and Department of Physics, Faculty of Science, University of Zagreb, Bijeni\v{c}ka cesta 32, HR-10000 Zagreb, Croatia \and Department of Physics, P.O. Box 64, FI-00014 University of Helsinki, Finland }

   \date{Received ; accepted}

\authorrunning{Miettinen}
\titlerunning{SABOCA, \textit{Herschel}, and \textit{WISE} views of G304.74+01.32}

\abstract{Filamentary molecular clouds, such as many of the infrared dark clouds (IRDCs), can undergo hierarchical 
fragmentation into substructures (clumps and cores) that can eventually collapse to form stars.}
{We aim to determine the occurrence of fragmentation into cores in the clumps of the filamentary IRDC G304.74+01.32 (hereafter, G304.74). We also aim to determine the basic physical characteristics (e.g. mass, density, and young stellar 
object (YSO) content) of the clumps and cores in G304.74.}
{We mapped the G304.74 filament at 350~$\mu$m using the SABOCA bolometer. The new SABOCA data have a factor of 2.2 times higher 
resolution than our previous LABOCA 870~$\mu$m map of the cloud ($9\arcsec$ versus $19\farcs86$). 
We also employed the \textit{Herschel} far-infrared (IR) and submillimetre, 
and \textit{Wide-field Infrared Survey Explorer} (\textit{WISE}) IR imaging data available for G304.74. 
The \textit{WISE} data allowed us to trace the IR emission of the YSOs associated with the cloud.}
{The SABOCA 350~$\mu$m data show that G304.74 is composed of a dense filamentary structure with a mean width of only $0.18\pm0.05$~pc. The percentage of LABOCA clumps that are found to be fragmented into SABOCA cores is $36\%\pm16\%$, but the irregular morphology of some of the cores suggests that this multiplicity fraction could be higher. The \textit{WISE} data suggest that 
$65\%\pm18\%$ of the SABOCA cores host YSOs. The mean dust temperature of the clumps, derived by comparing the \textit{Herschel} 250, 350, and 500~$\mu$m flux densities, was found to be $15.0 \pm 0.8$~K. The mean mass, beam-averaged H$_2$ column density, and H$_2$ number density of the LABOCA clumps are estimated to be $55\pm10$~M$_{\sun}$, $(2.0\pm0.2)\times10^{22}$~cm$^{-2}$, and $(3.1\pm0.2)\times10^4$~cm$^{-3}$. The corresponding values for the SABOCA cores are $29\pm3$~M$_{\sun}$, $(2.9\pm0.3)\times10^{22}$~cm$^{-2}$, and $(7.9\pm1.2)\times10^4$~cm$^{-3}$. The G304.74 filament is estimated to be thermally supercritical by a factor of $\gtrsim3.5$ on the scale probed by 
LABOCA, and by a factor of $ \gtrsim1.5$ for the SABOCA filament.}
{Our data strongly suggest that the IRDC G304.74 has undergone hierarchical fragmentation. On the scale where the clumps have 
fragmented into cores, the process can be explained in terms of gravitational Jeans instability. Besides the filament being fragmented, the finding of embedded YSOs in G304.74 indicates its thermally supercritical state, although the potential non-thermal (turbulent) motions can render the cloud a virial equilibrium system on scale traced by LABOCA. The IRDC G304.74 has a seahorse-like morphology in the \textit{Herschel} images, and the filament appears to be attached by elongated, perpendicular striations. This is potentially evidence that G304.74 is still accreting mass from the surrounding medium, and the accretion process can contribute to the dynamical evolution of the main filament. One of the clumps in G304.74, IRAS~13039-6108, is already known to be associated with high-mass star formation, but the remaining clumps and cores in this filament might preferentially form low and intermediate-mass stars owing to their mass reservoirs and sizes. Besides the presence of perpendicularly oriented, dusty striations and potential embedded intermediate-mass YSOs, G304.74 is a relatively nearby ($d\sim2.5$~kpc) IRDC, which makes it a useful target for future star formation studies. Owing to its observed morphology, we propose that G304.74 could be nicknamed the Seahorse Nebula.}


   \maketitle
%

\section{Introduction}

Mid-infrared (IR) satellite imaging surveys in the past showed that the Galactic plane 
exhibits dark absorption features against the Galactic mid-IR background radiation field 
(\cite{perault1996}; \cite{egan1998}; \cite{simon2006a}; \cite{peretto2009}).
Some of these features also show dust continuum emission at far-IR and 
(sub-)millimetre wavelengths, and hence are real, physical interstellar dust clouds, 
which are known as IR dark clouds or IRDCs (e.g. \cite{wilcock2012}). Observational studies of 
IRDCs have demonstrated that some of the clouds are able to give birth to high-mass 
($M_{\star}\gtrsim8$~M$_{\sun}$; spectral type B3 or earlier) stars 
(e.g. \cite{rathborne2006}; \cite{beuther2007}; \cite{chambers2009}; 
\cite{battersby2010}; \cite{zhang2011}). However, high-mass star 
formation occurs most likely in the most massive members of the IRDC population, 
while the majority of IRDCs might only give birth to low to intermediate-mass 
stars and stellar clusters (\cite{kauffmann2010}, hereafter KP10).

In this paper, we present a follow-up study of the IRDC G304.74+01.32 (hereafter, G304.74) that 
was first identified in the \textit{Midcourse Space Experiment} (\textit{MSX}) 
survey of the Galactic plane (\cite{simon2006a}). Beltr{\'a}n et al. (2006) 
mapped this IRDC in the 1.2~mm dust continuum emission using the SEST IMaging Bolometer Array (SIMBA) 
on the Swedish-ESO Submillimetre Telescope (SEST). Another dust continuum study of G304.74 was 
carried out by Miettinen \& Harju (2010; hereafter, Paper~I), who mapped the cloud at 870~$\mu$m using 
the Large APEX BOlometer CAmera (LABOCA; \cite{siringo2009}) on the Atacama Pathfinder EXperiment (APEX). In Paper~I, 
we identified 12 LABOCA clumps along the G304.74 filament.

The IRDC G304.74 does not lie within the area covered by the \textit{Spitzer} Space Telescope's IR survey 
images of the Galaxy (at the cloud's longitude, the \textit{Spitzer} images cover the 
Galactic latitudes $\left| b \right|\leq1\degr$). Hence, in Paper~I we had to rely 
on the \textit{MSX} and \textit{Infrared Astronomical Satellite} (\textit{IRAS}) images 
to examine how the LABOCA 870~$\mu$m emission of G304.74 compares with its appearance in the 
mid and far-IR bands. Of the 12 identified LABOCA clumps, four ($33\%\pm17\%$, where the
quoted uncertainty represents the Poisson error on counting statistics) were 
found to be associated with \textit{MSX} 8~$\mu$m sources. 
Three of these \textit{MSX}-bright clumps are associated with \textit{IRAS} 
point sources (the \textit{IRAS} sources 13037-6112, 13039-6108, and 
13042-6105; hereafter, IRAS~13037, etc.). The remaining eight clumps ($67\%\pm24\%$) were 
found to appear dark in the \textit{MSX} 8~$\mu$m image.

Miettinen (2012a; hereafter, Paper~II) presented the first molecular spectral line 
observations along the entire filamentary structure of G304.74. All 12 of the aforementioned
LABOCA 870~$\mu$m identified clumps were observed in C$^{17}$O$(2-1)$, and selected clumps
were also observed in $^{13}$CO$(2-1)$, SiO$(5-4)$, and CH$_3$OH$(5_k-4_k)$. 
The C$^{17}$O$(2-1)$ radial velocities of the observed clumps across the whole cloud were found 
to be comparable to each other, which shows that G304.74 is a spatially coherent 
filamentary structure (the average C$^{17}$O$(2-1)$ radial velocity of the filament was 
derived to be $\langle {\rm v}_{\rm LSR} \rangle=-26.7$~km~s$^{-1}$). In Paper~II, the kinematic distance of G304.74 
was derived to be $d=2.54\pm0.66$~kpc using the Galactic rotation curve of Reid et al. (2009). 
This is consistent with a value of 2.44~kpc derived earlier by Fontani 
et al. (2005) for IRAS~13039 using the Brand \& Blitz (1993) rotation curve. For comparison, the revised kinematic distance calculator of Reid et al. (2014) yields a distance of $d=2.31^{+0.72}_{-0.61}$~kpc for G304.74 (based on the model A5 of Reid et al. (2014); see their Table~4). The Bayesian distance calculator of Reid et al. (2016) gives a distance of $d=2.03\pm0.63$~kpc for G304.74 when no prior information is given to resolve the near-far kinematic distance ambiguity\footnote{The web-based Reid et al. (2014, 2016) kinematic distance calculators are available at \url{http://bessel.vlbi-astrometry.org}.}. The Bayesian approach of Reid et al. (2016) can significantly improve the accuracy and reliability of the distances to sources that are associated with the spiral arms of the Galaxy. However, no spiral arm association could be determined for G304.74, and hence its distance estimate is fairly uncertain. For this reason, the kinematic distance derived in Paper~II, which is consistent within the uncertainties with the other aforementioned values, is adopted in the present work.

The line observations presented in Paper~II were used 
to revise the distance-dependent physical properties of G304.74 and its clumps that were derived earlier 
in Paper~I, and to examine the gas kinematics, dynamical state, and some chemical 
properties of the clumps, such as the degree of CO depletion. The 870~$\mu$m dust-inferred total mass 
of the filament was estimated to be $\sim1\,200$~M$_{\sun}$, and its  
projected extent along the long axis was measured to be 8.9~pc. The corresponding line mass (mass per unit length) of the LABOCA filament is about 135~M$_{\sun}$~pc$^{-1}$. The LABOCA clumps 
were found to exhibit transonic to supersonic non-thermal motions, and a 
virial parameter analysis suggested that most of them ($75\%\pm25\%$) are gravitationally 
bound. The five $^{13}$CO$(2-1)$ lines detected towards G304.74 were found to exhibit blue asymmetric profiles, 
which was interpreted to be an indication of large-scale infall motions. However, 
the presence of different velocity components along the line of sight can 
also produce such line profiles (e.g.~\cite{henshaw2012}). No evidence 
of significant CO depletion in G304.74 was found in Paper~II, which can be understood if the 
feedback effects of star formation in the cloud, such as heating and protostellar outflows, have released 
the CO molecules from the icy grain mantles back into the gas phase. We note that Fontani et al. (2012), 
who used C$^{18}$O$(3-2)$ observations, derived a very high CO depletion factor of $f_{\rm D}=22$ towards 
a selected position in G304.74 (their source 13039-6108c6). However, their target position appears to lie outside significant LABOCA 870~$\mu$m emission of the cloud (just outside the $3\sigma$ level, and $17\farcs2$ to the south-east of the peak position of our LABOCA clump called SMM~9; see Sect.~2.2). Indeed, no strong C$^{18}$O$(3-2)$ emission is to be expected from the target position observed by Fontani et al. (2012) because 
the critical density of the transition is $3\times10^4$~cm$^{-3}$ at 10~K. For comparison, in Paper~II we found 
no evidence of CO depletion towards SMM~9 ($f_{\rm D}=1.2\pm0.6$), which was the nearest of our target positions to that observed by Fontani et al. (2012). 

Regarding the clumps' ability to form massive stars, it was found in Paper~II that none of the clumps fulfils the mass-radius threshold for 
massive star formation proposed by KP10. Hence, it was
concluded that the clumps of G304.74 might only be able to give birth to low to intermediate-mass stars, 
but owing to the high masses of the clumps, the star formation process could occur in a clustered mode.

Regarding the global characteristics of G304.74, in Paper~II it was found that the LABOCA filament 
as a whole has the potential to be close to virial equilibrium between gravitational and turbulent energies. 
Our analysis of the cloud's virial line mass suggested that it could receive additional support against 
radial collapse from a poloidal magnetic field component (following the models of Fiege 
\& Pudritz (2000a)). We also suggested that the ambient 
turbulent ram pressure might be an important factor in the dynamics of G304.74, which is 
qualitatively consistent with a scenario where the filament was formed 
by converging, supersonic turbulent flows (Paper~I). 
The projected separations between the LABOCA clumps were found to be consistent with a theoretical prediction 
of the so-called sausage-type fluid instability, which can lead to a periodic fragmentation of the filament (e.g. \cite{chandrasekhar1953}), and which has been found to be a potential fragmentation mechanism for other IRDC filaments as well (e.g. \cite{jackson2010}; \cite{miettinen2012b}; \cite{busquet2013}). 

In this paper, we present the results of our new dust continuum follow-up imaging of the IRDC G304.74, which was carried out using the Submillimetre APEX Bolometer Camera (SABOCA; \cite{siringo2010}) at 350~$\mu$m. In particular, the 
present SABOCA data allowed us to build a new, $\sim0.1$~pc resolution view of 
the dense internal structure of the cloud and its 
fragmentation into star-forming units. To complement our SABOCA data, 
we employed our previous LABOCA data, and the far-IR and submm data from \textit{Herschel} (\cite{pilbratt2010})\footnote{\textit{Herschel} is an ESA space observatory with science instruments provided
by European-led Principal Investigator consortia and with important
participation from NASA.}. We also used the observations by the 
\textit{Wide-field Infrared Survey Explorer} (\textit{WISE}; \cite{wright2010}), which provide a sensitive, high-resolution view of the near and mid-IR
properties of G304.74. Most notably, the \textit{WISE} data sets allowed us to characterise the star-forming activity in G304.74. As mentioned above, G304.74 is not covered by the \textit{Spitzer} IR observations, and hence 
the \textit{WISE} data for the cloud are particularly useful.

The layout of this paper is as follows. The SABOCA observations, the data 
reduction, and the ancillary data sets are described in Sect.~2. The analysis 
and its results are presented in Sect.~3. In Sect.~4, we discuss the results, 
and in Sect.~5 we summarise our results and present our main conclusions.

\section{Observations, data reduction, and ancillary data}

\subsection{SABOCA 350~$\mu$m imaging}

A field of $12\farcm5 \times 14\farcm2$ (0.05~deg$^2$), centred on G304.74, 
was mapped with SABOCA (project {\tt 089.F-9310(A)}; PI: O.~Miettinen) on the APEX 12~m telescope 
(\cite{gusten2006}). The instrument is a 37-channel on-sky transition edge sensor (TES) bolometer array operating at 350~$\mu$m, with a nominal angular resolution of $7\farcs7$ (half power beam width (HPBW)). The effective field of view of the array is $1\farcm5$. The SABOCA passband
has an equivalent width of about 120~GHz centred on an effective frequency 
of 852~GHz. The observations were carried out on 6~June 2012. The average scanning speed was 
$\sim1\farcm4$~s$^{-1}$, and the total on-source integration time was about 
3 h (during the UTC time range 02:01--05:16). 

The atmospheric attenuation at 350~$\mu$m was monitored using the sky-dip method, and the 
zenith opacity was found to be in the range 
$\tau_{\rm z}^{350\,\mu {\rm m}}=1.09-1.22$. The amount of precipitable water 
vapour was $\sim0.5$~mm. The telescope focus and pointing were optimised 
and checked at regular intervals on the pla\-nets Mars and Saturn, the galaxy 
Centaurus A (NGC~5128), the \ion{H}{ii} region and OH maser source 
034.257+0.155, and the massive young stellar object (YSO) and H$_2$O maser 
source G305.80-0.24 (B13134). The absolute calibration uncertainty for SABOCA is $\sim30\%$ (\cite{dumke2010}). 

The SABOCA data were reduced using the Comprehensive Reduction Utility for 
SHARC-2 ({\tt CRUSH-2}, version 2.12-2; \cite{kovacs2008})\footnote{{\tt CRUSH-2} is publicly available, and can be retrieved at \url{http://www.submm.caltech.edu/~sharc/crush/index.htm.}}. We used the pipeline 
iterations with the default reduction parameters, and also with the 
{\tt extended} and {\tt deep} reduction options, which better preserve the 
extended source structures and recover the point-like sources, respectively. 
The {\tt deep} option, which performs strong spatial filtering (angular scales 
larger than $37\farcs5$ are not recovered), turned out to give the best image 
quality, and hence was chosen for the subsequent analysis. 
The default settings and {\tt extended} option both resulted in maps with higher levels of noise than the {\tt deep} option. In addition, the {\tt extended} option led to artificial large-scale features at the 
edges of the map. 

To improve the image quality further, and increase the 
signal-to-noise (S/N) ratio of the detected emission, 
the map was smoothed with a Gaussian kernel of 
the size $4\farcs97$ (in addition to the instrument beam full 
width at half maximum (FWHM) of $7\farcs5$ assumed by {\tt CRUSH-2}). 
Hence, the effective angular resolution of our final 
SABOCA map is $9\arcsec$ (0.11~pc at 2.54~kpc). 
The gridding was performed with a cell size of $1\farcs5$. 
The resulting $1\sigma$ root-mean-square (rms) noise level in the final 
map, which is shown in the left panel in Fig.~\ref{figure:submm}, 
is $\sim200$~mJy~beam$^{-1}$. 

The SABOCA 350~$\mu$m imaging of G304.74 reveals its narrow, integral-shaped 
filamentary structure. The emission at low spatial frequencies, that is emission 
on large angular scales, was efficiently filtered out during the atmospheric 
sky noise removal. As is often the case in dust continuum maps observed with ground-based bolometer arrays, 
there are artificial bowls of negative emission, or holes around regions of bright 
emission. Besides the calibration uncertainties, the negative regions are likely to introduce uncertainties in the absolute source 
flux densities.

\subsection{LABOCA 870~$\mu$m data}

\begin{figure*}
\begin{center}
\includegraphics[scale=0.44]{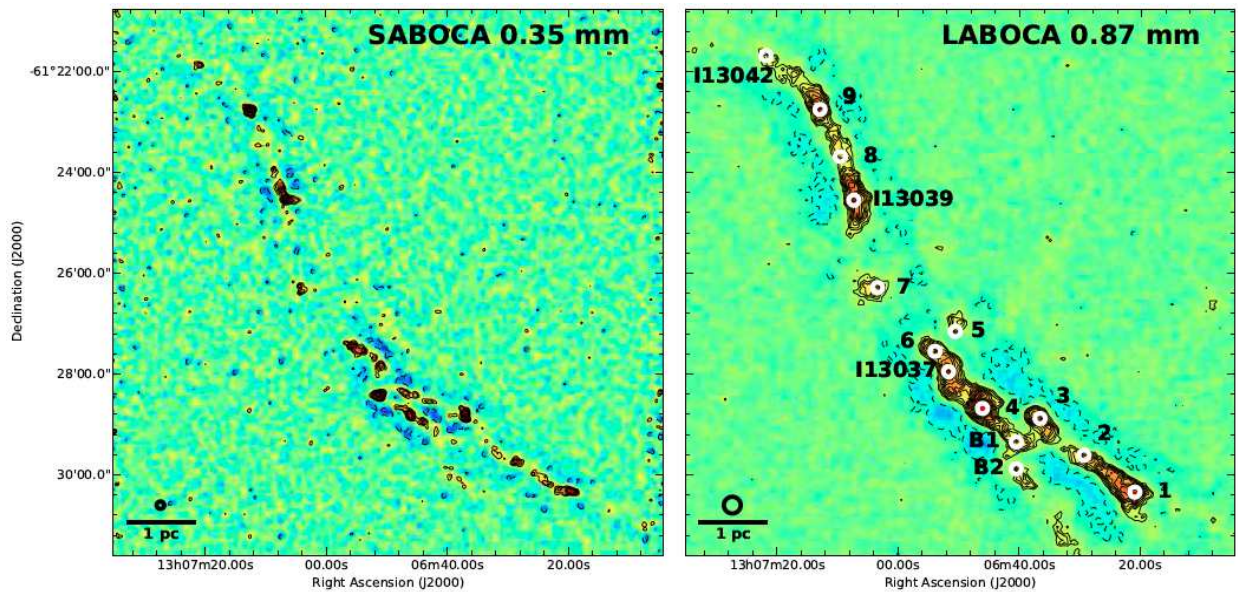}
\caption{SABOCA 350~$\mu$m (left) and LABOCA 870~$\mu$m (right) images of G304.74. In both panels, 
the colour scale is linear. The contours levels start from $3\sigma$, and progress 
in steps of $1\sigma$. The dashed contours show the negative features at the $-3\sigma$ level. The identified clumps are labelled on the LABOCA map, where the numbers refer to the SMM IDs (e.g. 1 means SMM~1), while the sources I13037, I13039, and I13042 are the three \textit{IRAS} sources in the filament. The white circles indicate the LABOCA peak positions of the clumps (see Sect.~3.1). The identified SABOCA sources are not labelled for legibility purposes (see Sect.~3.2), but they are indicated in the zoom-in images in Fig.~\ref{figure:fragments}. 
A scale bar of 1~pc projected length, and the effective beam size ($9\arcsec$ for SABOCA, $19\farcs86$ for LABOCA) are shown in the bottom left corner.}
\label{figure:submm}
\end{center}
\end{figure*}

Our LABOCA 870~$\mu$m observations were first  
published in Paper~I. The data were originally reduced using the BOlometer 
array data Analysis software ({\tt BoA}; \cite{schuller2012}). To be better 
comparable with the present SABOCA data, we re-reduced the LABOCA 
data using {\tt CRUSH-2} with the aforementioned {\tt deep} option. The data 
were smoothed using a Gaussian kernel of the size $3\farcs76$, 
while the instrument beam FWHM used in {\tt CRUSH-2} is assumed to be 
$19\farcs5$. Hence, the angular resolution of the final image is 
$19\farcs86$ (0.24~pc). The gridding was performed with a 
cell size of $4\farcs0$. The resulting $1\sigma$ rms noise level 
in the final map, which is presented in the right panel 
in Fig.~\ref{figure:submm}, is 40~mJy~beam$^{-1}$. 
For comparison, the LABOCA map presented in Paper~I 
had an angular resolution and rms noise of $20\farcs1$ and 30~mJy~beam$^{-1}$, 
respectively. 

The peak positions of the clumps in our revised LABOCA map were found to be systematically shifted to the east or south-east with respect to the map presented in Paper~I. On average, the angular offset was found to be $10\arcsec$ (median $9\farcs3$). Such angular offset between APEX continuum maps reduced using {\tt BoA} and {\tt CRUSH-2} is consistent with our previous finding for the Orion B9 star-forming region (\cite{miettinenetal2012}). However, the peak surface brightnesses of the clumps in the new LABOCA map were found to be in very good agreement with those derived in Paper~I, both the mean and median ratio between the two being 1.09. 

We note that similar negative features are seen around the LABOCA filament as in the SABOCA image, and hence some of the LABOCA flux densities can suffer from an uncertain absolute intensity scale. Besides this effect, another possible caveat is that our 870~$\mu$m continuum data could be contaminated by the $^{12}$CO$(3-2)$ rotational line emission, which has the rest frequency of 345~GHz (\cite{drabek2012}). The presence of significant CO$(3-2)$ emission is to be expected on the basis of the finding that CO does not appear to be depleted in G304.74 (Paper~II), but to properly quantify the CO contribution to the LABOCA signal would require the cloud to be mapped in 
CO$(3-2)$ emission.

\subsection{\textit{Herschel} far-infrared and submillimetre data}

The IRDC G304.74 lies within the Coalsack 
dark cloud field that was observed as part of the \textit{Herschel} Gould 
Belt Survey (GBS; \cite{andre2010})\footnote{For more details, 
see \url{http://gouldbelt-herschel.cea.fr}.}. To complement our APEX submm dust continuum data for G304.74, we downloaded the Spectral and Photometric Imaging Receiver (SPIRE; \cite{griffin2010}) 250~$\mu$m, 350~$\mu$m, and 500~$\mu$m images of G304.74 from the \textit{Herschel} Science Archive (HSA)\footnote{\url{herschel.esac.esa.int/Science_Archive.shtml}.}.

\begin{figure}[!htb]
\centering
\resizebox{0.9\hsize}{!}{\includegraphics{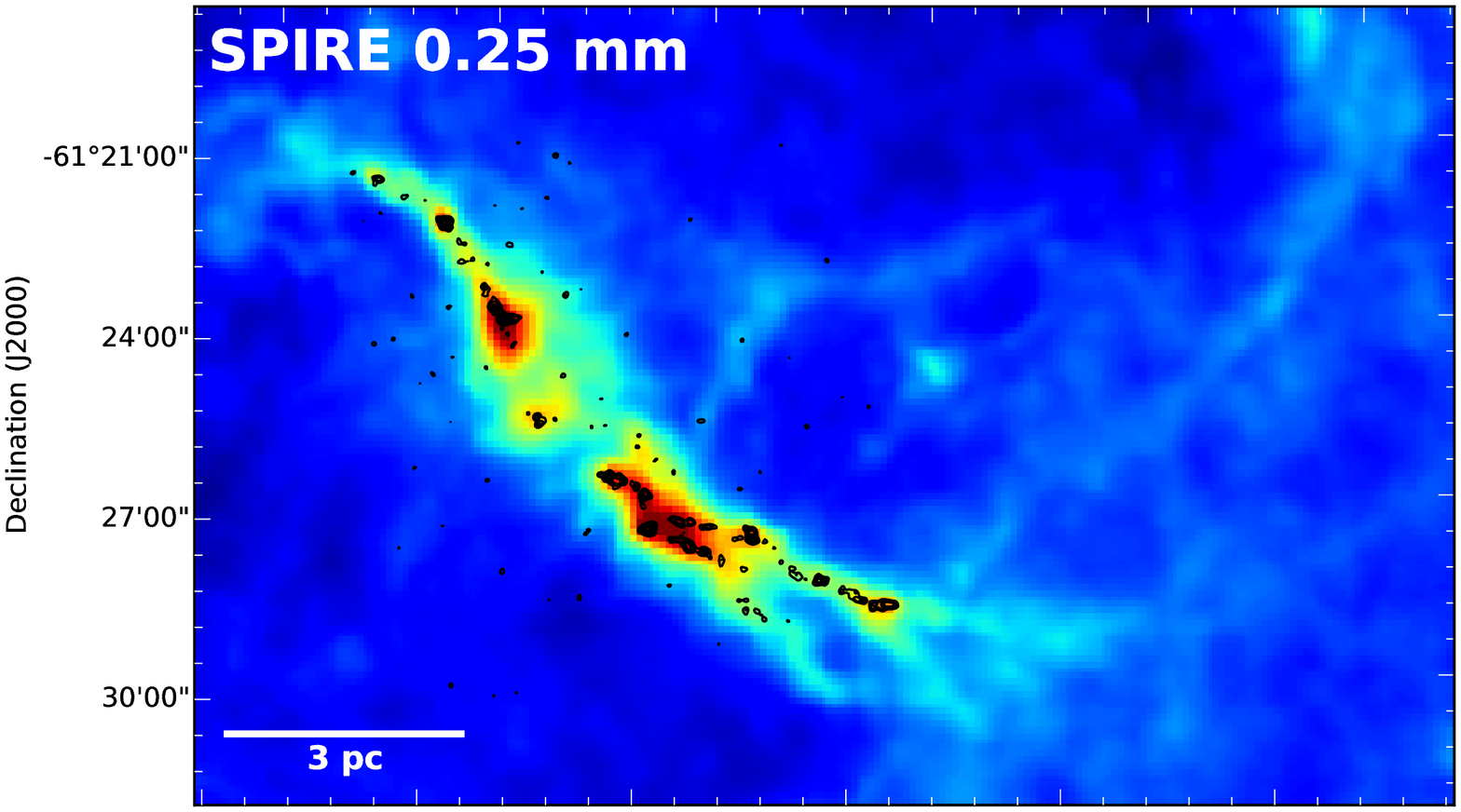}}\\[-5pt]
\resizebox{0.9\hsize}{!}{\includegraphics{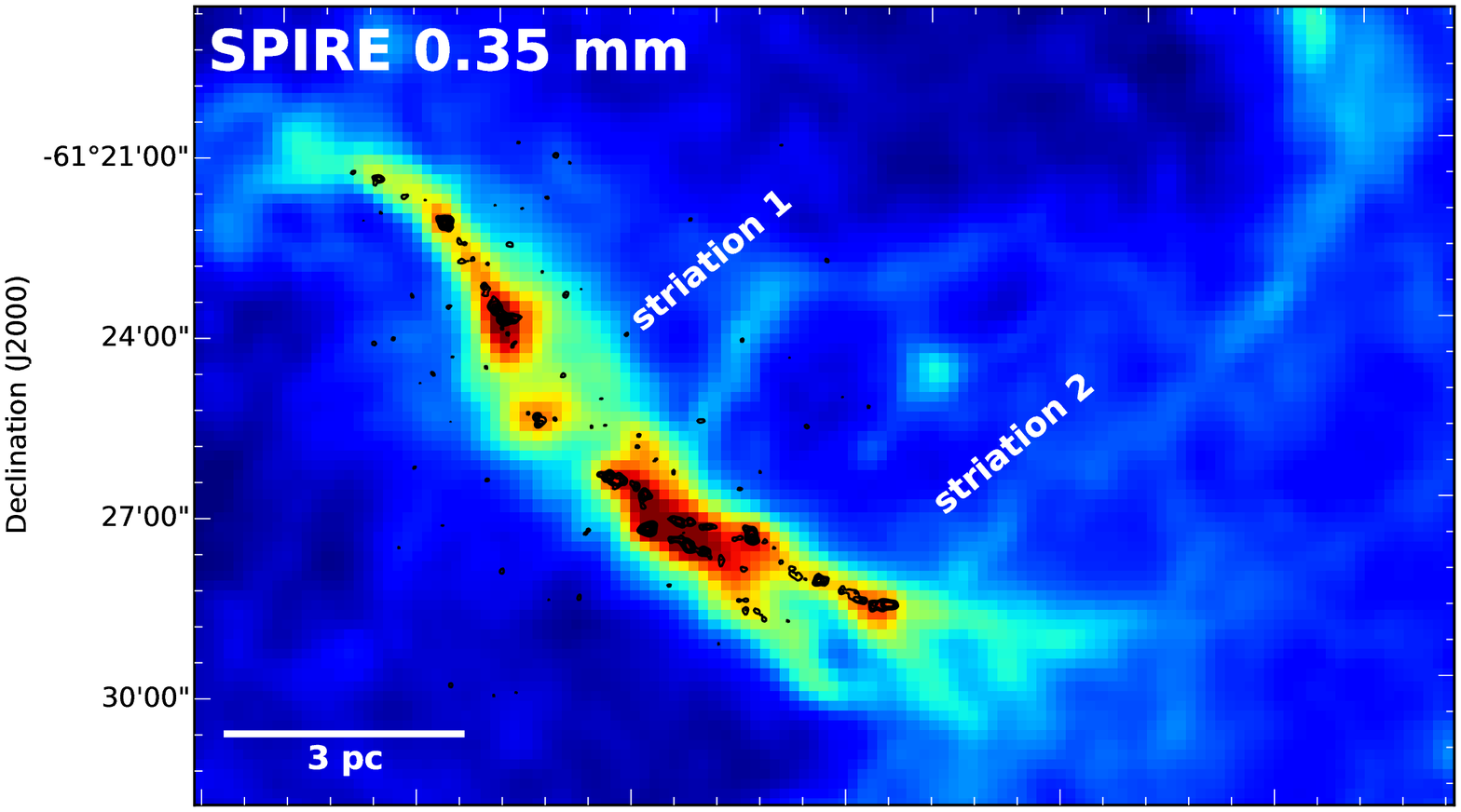}}\\[-5pt]
\resizebox{0.9\hsize}{!}{\includegraphics{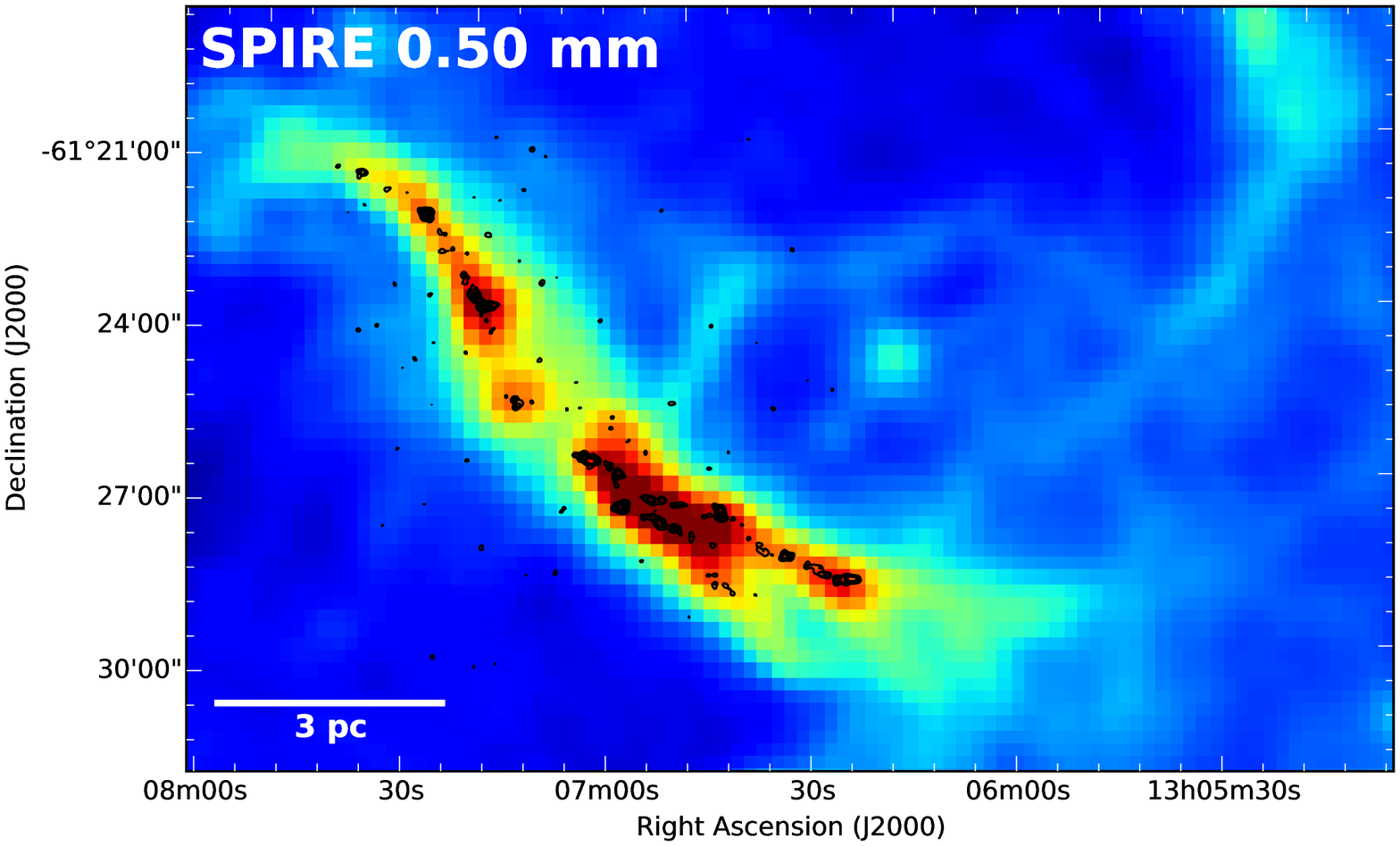}}\\[-5pt]
\caption{\textit{Herschel}/SPIRE images of G304.74 and its surroundings. The images 
are displayed in logarithmic colour-scale to improve the appearance of the 
low surface brightness features. The overlaid black contours show the SABOCA 
350~$\mu$m dust continuum emission as in Fig.~\ref{figure:submm}. A scale bar of 3~pc projected length is shown in the bottom left corner of each panel. The two obvious striations are labelled in the middle panel (striation~1 and striation~2).}
\label{figure:spire}
\end{figure}

The SPIRE images showing the IRDC G304.74 and its surroundings are shown in Fig.~\ref{figure:spire}. The angular resolutions of the maps are $18\farcs2$, $24\farcs9$, and $36\farcs3$ at 250~$\mu$m, 350~$\mu$m, and 500~$\mu$m, and the corresponding pixel sizes in the images are $6\arcsec$, $10\arcsec$, and $14\arcsec$, respectively. We note that the angular resolution of our SABOCA map is about 2.8 times higher than that of the SPIRE 350~$\mu$m map. On the other hand, the spatial filtering by SABOCA is stronger than in the \textit{Herschel} data, which makes SABOCA more sensitive to compact point-like sources, but insensitive to diffuse, large-scale emission seen by \textit{Herschel}. 

The projected morphology of G304.74 in the SPIRE images resembles that of a seahorse. Hence, following the nickname for the IRDC G338.4-0.4, that is the Nessie Nebula (\cite{jackson2010}), and that for the IRDC G11.11-0.11, the Snake Nebula (e.g. \cite{wang2014}, and references therein), we propose that G304.74 could be dubbed the Seahorse Nebula.

\subsection{\textit{WISE} near and mid-infrared data}

To reach a better understanding of the star formation activity in G304.74, we employed the 
IR data taken by \textit{WISE}. As mentioned in Sect.~1, the \textit{WISE} IR observations are of particular interest for G304.74, because the IRDC is not covered by the \textit{Spitzer} images owing to the high Galactic latitude of the cloud.

\begin{figure*}
\begin{center}
\includegraphics[scale=0.44]{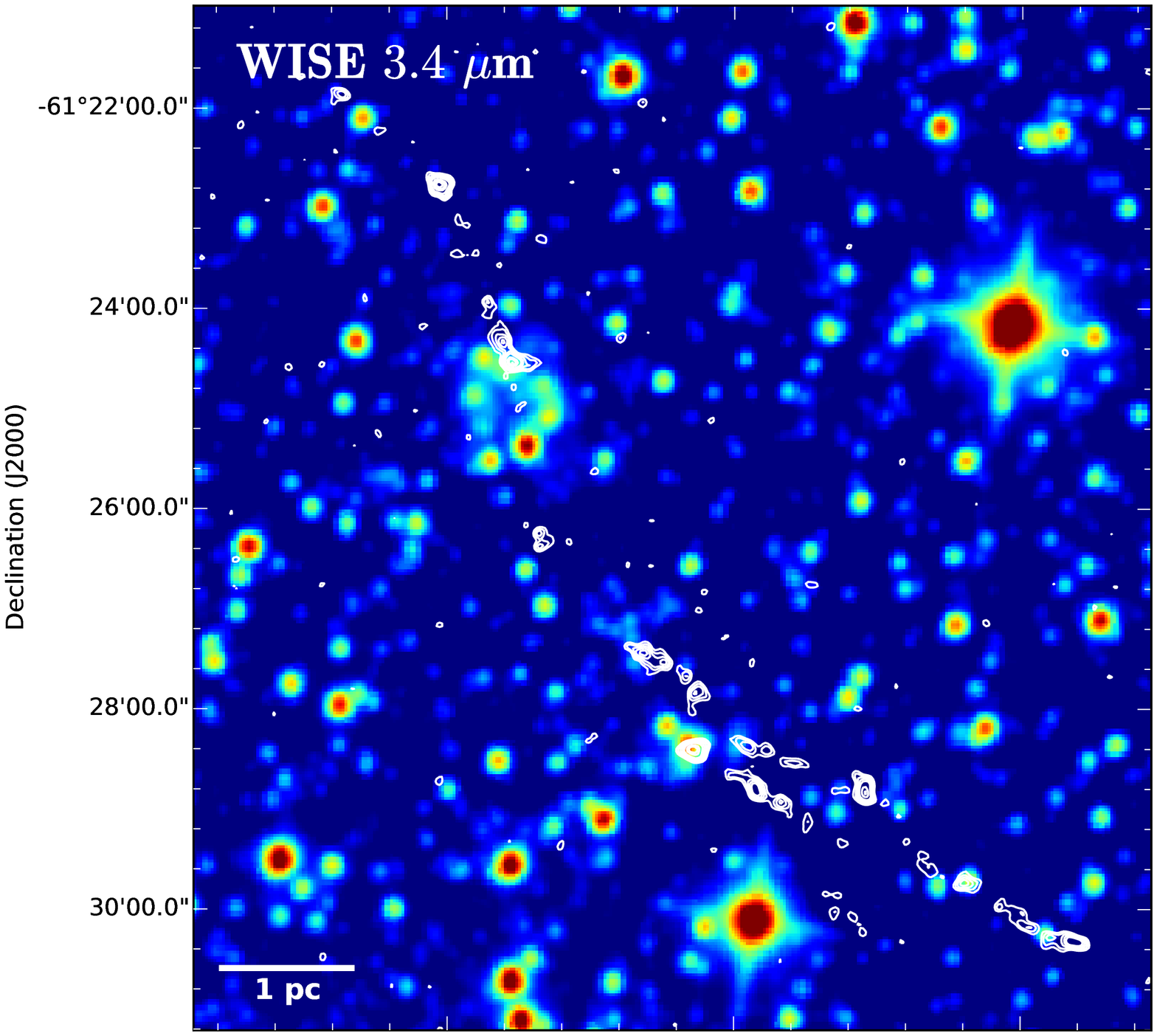}
\includegraphics[scale=0.44]{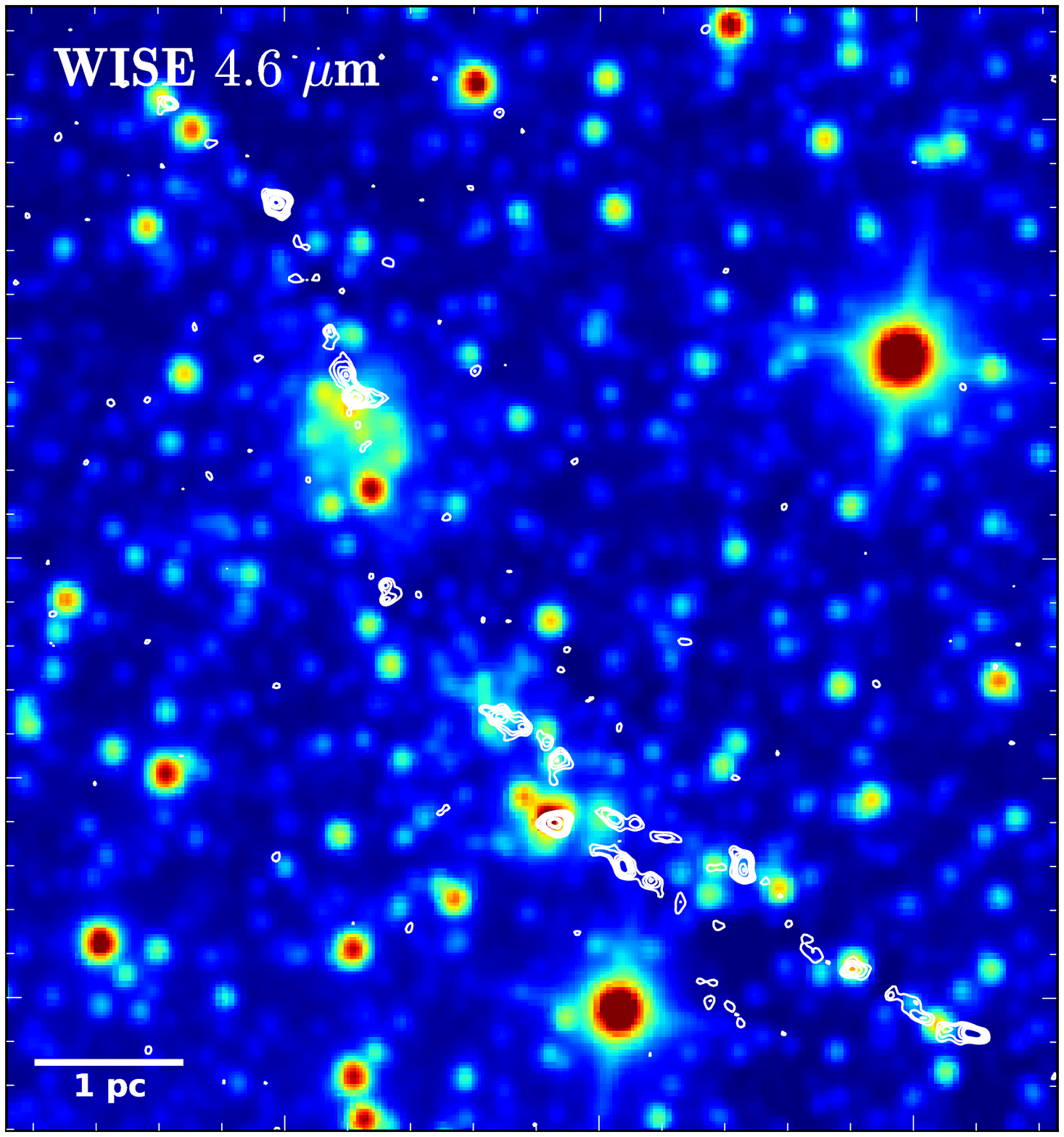}
\includegraphics[scale=0.44]{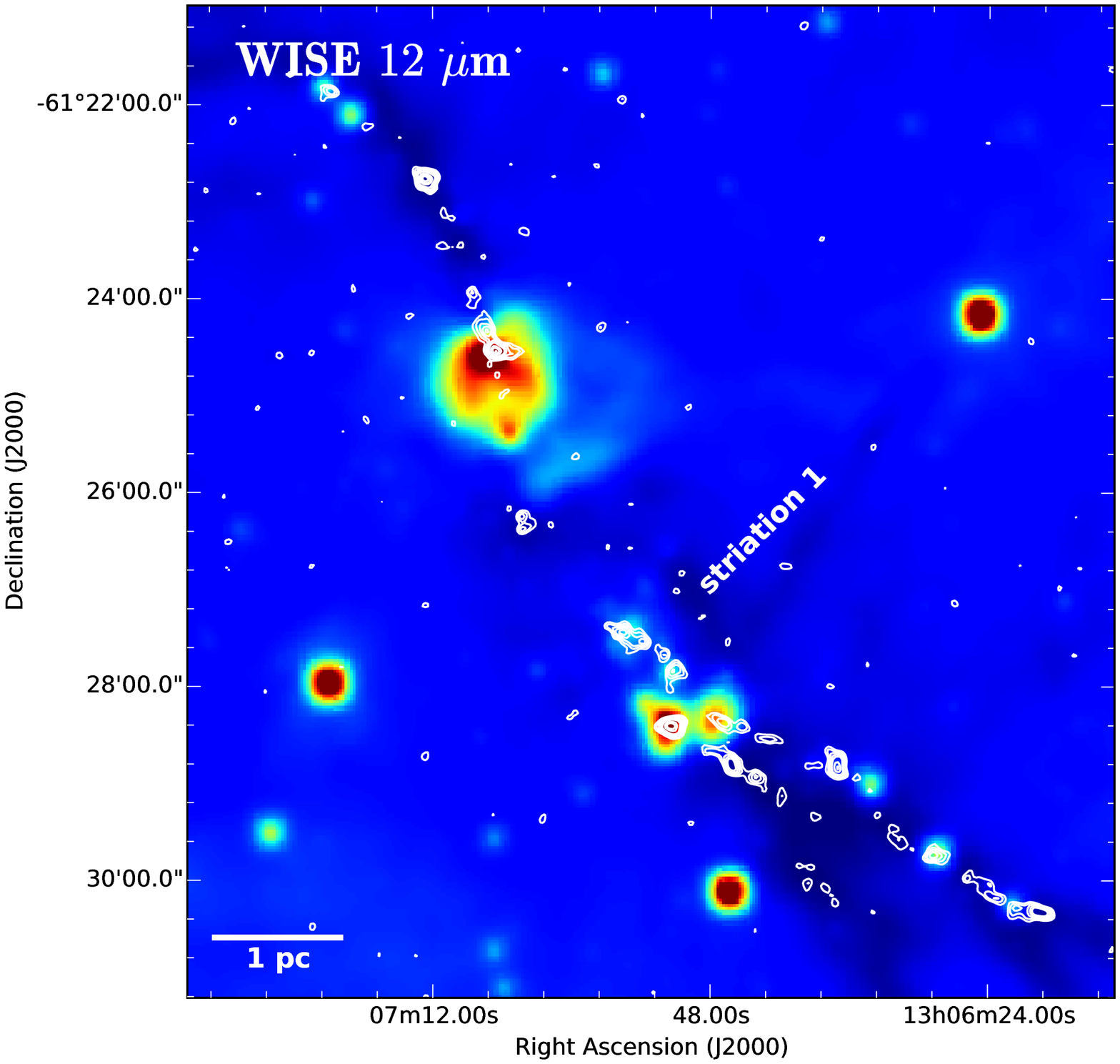}
\includegraphics[scale=0.44]{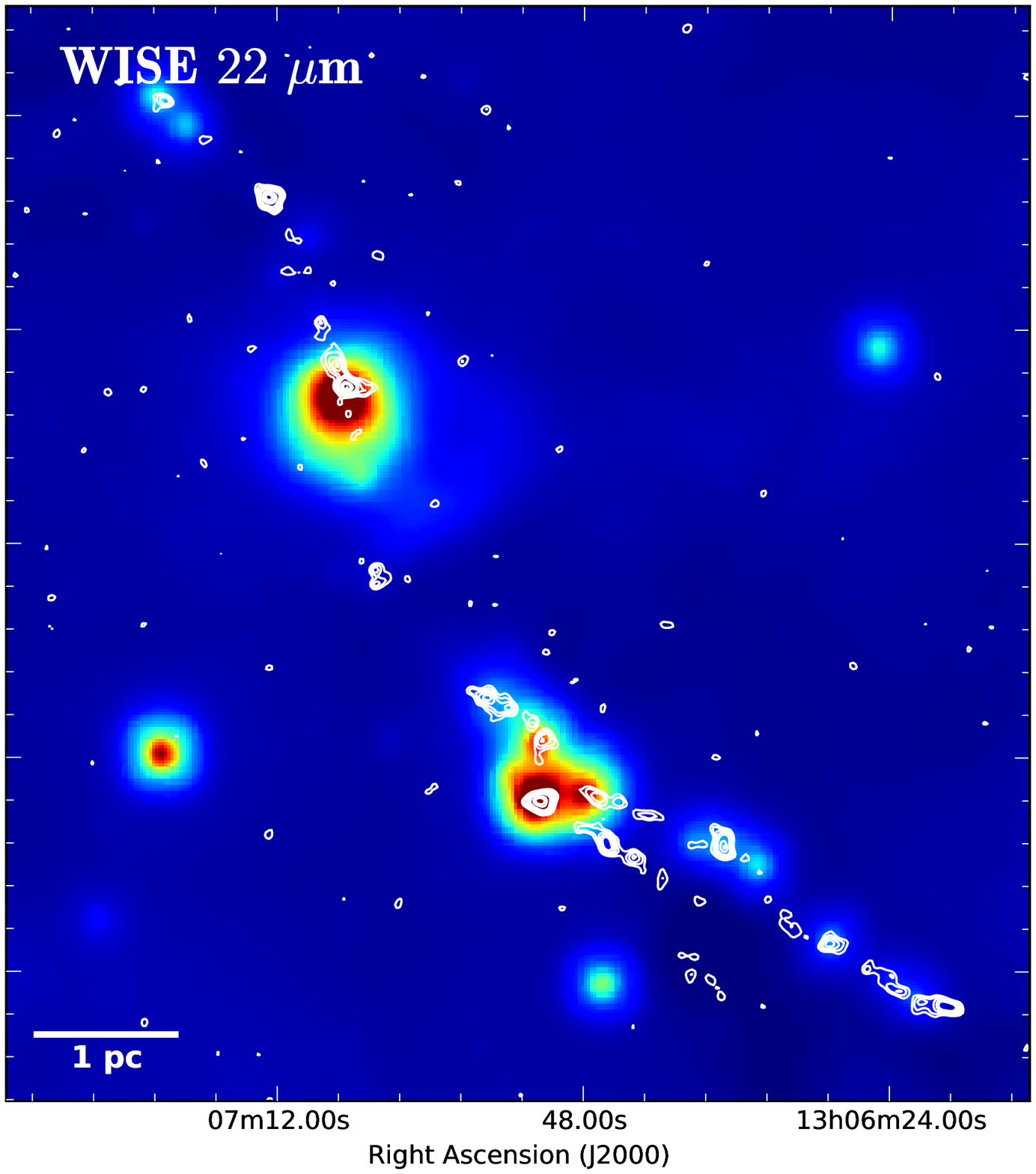}
\caption{\textit{WISE} images towards G304.74 overlaid with contours of SABOCA 
350~$\mu$m dust continuum emission from Fig.~\ref{figure:submm}. 
The images are shown with logarithmic scaling to improve the colour contrast. 
A scale bar of 1~pc projected length is 
shown in the bottom left corner of each panel. The striation~1 is labelled in the 12~$\mu$m image, while striation~2 lies outside the image boundary (cf.~Fig.~\ref{figure:spire}).}
\label{figure:wise}
\end{center}
\end{figure*}

The \textit{WISE} satellite surveyed the entire sky at 3.4, 4.6, 12, 
and 22~$\mu$m (channels W1--W4). The estimated $5\sigma$ point-source 
sensitivities at these wavelengths are better than 
0.068, 0.098, 0.86, and 5.4~mJy (or 16.83, 15.60, 11.32, and 8.0 in Vega magnitudes), where the exact value depends on the ecliptic 
latitude of the observed field. The angular resolution (FWHM) of the \textit{WISE} images 
in the aforementioned four bands is $6\farcs1$, $6\farcs4$, $6\farcs5$, and 
$12\farcs0$, respectively. The \textit{WISE} All-Sky Data Release is available through the 
NASA/IPAC Infrared Science Archive (IRSA)\footnote{\url{http://wise2.ipac.caltech.edu/docs/release/allsky/}.}. 
The \textit{WISE} images towards G304.74 are shown in Fig.~\ref{figure:wise}.

\section{Analysis and results}

\subsection{Source extraction from the SABOCA and LABOCA maps} 

To extract the sources from our new SABOCA 
350~$\mu$m map, we used the {\tt BLOBCAT} source extraction software, which is written in {\tt Python} (\cite{hales2012}). The code makes use of the the flood fill algorithm
to detect individual islands of pixels (or blobs) in two-dimensional images, and its 
performance was benchmarked against standard Gaussian fitting. While {\tt BLOBCAT} is most often used in extragalactic studies (e.g. \cite{hales2014}; \cite{smolcic2017}), the present work demonstrates its application in a Galactic study. 
As the background rms noise level, we used a uniform value of 
$0.2$~Jy~beam$^{-1}$. The S/N threshold for blob detection was set to 
${\rm S/N}_{\rm thres}=4$, and the flooding cut-off threshold was set to ${\rm S/N}_{\rm cut}=3$, that is the flooding was performed down to $3\sigma$ (for details, see Sect.~2.2 in \cite{hales2012}). 

Similarly, and to be consistent with the SABOCA source extraction, the sources in the LABOCA map were extracted using {\tt BLOBCAT}. Again, a uniform rms noise level (40~mJy~beam$^{-1}$) was used in the extraction process. Compared to the SABOCA image, G304.74 appears as a more continuous filamentary structure in the LABOCA map. Hence, to recover the visually prominent clumps (Fig.~\ref{figure:submm}, right panel), which were also identified in Paper~I, the values of ${\rm S/N}_{\rm thres}$ and ${\rm S/N}_{\rm cut}$ were both varied from 3 to 7.5, with the most succesful extraction (five clumps) yielded by the parameters ${\rm S/N}_{\rm thres}={\rm S/N}_{\rm cut}=5$. We note that varying values of ${\rm S/N}_{\rm thres}$ and ${\rm S/N}_{\rm cut}$ are not expected to introduce any significant bias in the blob detection from the G304.74 LABOCA filament owing to its fairly continuous structure, and the fact that the clumps are of different surface brightness. 

The identified LABOCA clumps are tabulated in Table~\ref{table:LABOCA}. 
Besides the 12 clumps identified in Paper~I using the two-dimensional contouring 
algorithm {\tt clfind2d} (\cite{williams1994}), two additional sources were extracted by the present {\tt BLOBCAT} analysis. We call these sources BLOB~1 and BLOB~2, where the former lies in between SMM~3 and SMM~4, and the latter is a separate blob south of BLOB~1 (Fig.~\ref{figure:submm}, right panel). In Table~\ref{table:LABOCA}, we list the positions of the peak emission of the LABOCA clumps, the detection S/N ratio, the peak surface brightness, the integrated flux density, and the effective radius of the source. The aforementioned clump porperties were output by {\tt BLOBCAT}, and the intensity and flux density uncertainties were calculated by taking the absolute calibration uncertainty into account, which was added in quadrature to the uncertainty output by {\tt BLOBCAT}. The effective radius tabulated in Table~\ref{table:LABOCA} is defined as $R_{\rm eff}=\sqrt{A_{\rm proj}/\pi}$, where $A_{\rm proj}$ is the projected area within the source, and it was calculated from the total number of pixels assigned to the clump by {\tt BLOBCAT}. The clumps SMM~6, SMM~8, IRAS~13042, and BLOB~1 were found to have an effective radius smaller than half the beam FWHM. Hence, an 
upper limit of $<0.5\times \theta_{\rm beam}$ was assigned to the radius of these sources. The effective radii 
were not corrected for the beam size. We note that {\tt BLOBCAT} does not provide parametric source
sizes for a quantitative analysis. 

The identified SABOCA sources are tabulated in Table~\ref{table:SABOCA}. Owing to the small sizes of the sources, we call them cores, as opposed to the larger clumps detected in the LABOCA map. The LABOCA clumps SMM~1, SMM~4, SMM~6, and IRAS~13039 were resolved into two or more SABOCA cores (see Fig.~\ref{figure:fragments}). We call these cores as SMM~1a, SMM~1b, etc., where the alphabetical tags follow the order of an increasing S/N ratio of the source. The LABOCA sources SMM~5, SMM~8, and BLOB~1 were not detected in our SABOCA map. The reason for this is that the emission was resolved out at $9\arcsec$ resolution. For example, the latter sources were recovered in the SABOCA map after smoothing it to the resolution of our LABOCA map ($19\farcs86$), where the corresponding $1\sigma$ rms noise level is 180~mJy~beam$^{-1}$.

Considering the cloud physical properties that depend on the flux density of the submm dust emission, such as the dust-based mass estimates, it is worth mentioning that some 
percentage of the emission is being filtered out in our bolometer dust emission maps (Sect.~2.1). By comparing the SABOCA 350~$\mu$m flux density of the G304.74 filament to that estimated from the \textit{Herschel} 350~$\mu$m map suggests that as much as $\sim80\%$ of the lower surface brightness 350~$\mu$m emission might have been filtered out in the present SABOCA map. However, this should be taken as a rough estimate only because the \textit{Herschel}/SPIRE images used in the present study have not been calibrated to an absolute intensity scale. Similarly, to estimate how much flux is filtered out in our LABOCA 870~$\mu$m map, we used the \textit{Planck} 353~GHz thermal dust emission map as a reference (a public data release~2 product available through IRSA; see \cite{planck2014}). To take the frequency difference between our LABOCA data and the \textit{Planck} data into account, the \textit{Planck} flux was scaled down by assuming that the dust emissivity index is $\beta = 1.8$, which corresponds to a Galactic mean value derived through \textit{Planck} observations (\cite{planck2011}). This way, we estimate that about $\sim35\%$ of the 870~$\mu$m emission might have been filtered out in our LABOCA map. The filtered dust emission should be kept in mind in the calculation of the dust-based physical properties of G304.74. 

A quantitative estimate of the completeness of our census of the SABOCA and LABOCA sources in G304.74 could be obtained through simulations of artificial source extractions (see e.g. \cite{weiss2009}; \cite{konyves2015}; \cite{marsh2016}). However, rather than being a general (blind) dust continuum imaging survey, the present study focuses on a continuous high-column density filament. The separation of such filament into clumps and cores can be highly sensitive to the source extraction algorithm used, and the adopted extraction parameter settings. This, together with the fact that the background SABOCA and LABOCA emissions around the G304.74 filament show fluctuations (the negative features in Fig.~\ref{figure:submm}), renders the completeness analysis problematic. On the other hand, our SABOCA and LABOCA sources are of high significance (${\rm S/N}_{\rm SABOCA}=4.0-11.8$ with an average of $\langle {\rm S/N}_{\rm SABOCA}\rangle=7.2$; ${\rm S/N}_{\rm LABOCA}=5.8-14.6$ with an average of $\langle {\rm S/N}_{\rm LABOCA}\rangle=9.6$). Hence, our submm source catalogues are unlikely to contain spurious sources.

\begin{table*}
\caption{LABOCA 870 $\mu$m clumps in G304.74.}
\begin{minipage}{2\columnwidth}
\centering
\renewcommand{\footnoterule}{}
\label{table:LABOCA}
\begin{tabular}{c c c c c c c c}
\hline\hline 
Source & $\alpha_{2000.0}$ & $\delta_{2000.0}$ & S/N & $I_{870}^{\rm peak}$ & $S_{870}$ & \multicolumn{2}{c}{$R_{\rm eff}$}\\
       & [h:m:s] & [$\degr$:$\arcmin$:$\arcsec$] & & [Jy~beam$^{-1}$] & [Jy] & [\arcsec] & [pc]\\
\hline
SMM 1 & 13 06 20.89 & -61 30 20.51 & 12.1 & $0.48\pm0.07$ & $1.53\pm0.18$ & 25.1 & $0.31\pm0.08$ \\
SMM 2 & 13 06 29.29 & -61 29 36.67 & 9.1 & $0.36\pm0.05$ & $0.63\pm0.06$ & 17.0 & $0.21\pm0.05$ \\
SMM 3 & 13 06 36.56 & -61 28 52.79 & 11.1 & $0.44\pm0.07$ & $0.82\pm0.11$ & 15.8 & $0.19\pm0.05$\\
SMM 4 & 13 06 46.06 & -61 28 40.90 & 14.6 & $0.58\pm0.08$ & $1.47\pm0.17$ & 18.5 & $0.23\pm0.06$ \\
IRAS 13037-6112 & 13 06 51.65 & -61 27 56.95 & 11.5 & $0.46\pm0.07$ & $1.24\pm0.14$ & 14.5 & $0.18\pm0.05$\\
SMM 5 & 13 06 50.54 & -61 27 08.94 & 6.2 & $0.25\pm0.05$ & $0.28\pm0.05$ & 11.7 & $0.14\pm0.04$ \\
SMM 6 & 13 06 53.88 & -61 27 32.96 & 10.3 & $0.41\pm0.06$ & $0.41\pm0.06$ & $<9.9$ & $<0.15$ \\
SMM 7 & 13 07 03.38 & -61 26 17.00 & 8.1 & $0.33\pm0.05$ & $0.52\pm0.07$ & 14.0 & $0.17\pm0.04$ \\
IRAS 13039-6108 & 13 07 07.28 & -61 24 33.00 & 12.8 & $0.51\pm0.07$ & $1.41\pm0.16$ & 17.4 & $0.21\pm0.06$ \\
SMM 8 & 13 07 09.51 & -61 23 41.00 & 7.0 & $0.28\pm0.05$ & $0.28\pm0.05$ & $<9.9$ & $<0.15$\\
SMM 9 & 13 07 12.85 & -61 22 44.99 & 11.2 & $0.45\pm0.07$ & $0.83\pm0.10$ & 13.0 & $0.16\pm0.04$ \\
IRAS 13042-6105 & 13 07 21.74 & -61 21 40.95 & 5.8 & $0.23\pm0.05$ & $0.23\pm0.05$ & $<9.9$ & $<0.15$\\
\hline 
BLOB 1\tablefootmark{a} & 13 06 40.46 & -61 29 20.84 & 8.2 & $0.33\pm0.05$ & $0.33\pm0.05$ & $<9.9$ & $<0.15$\\
BLOB 2\tablefootmark{a} & 13 06 40.44 & -61 29 52.84 & 6.8 & $0.27\pm0.05$ & $0.31\pm0.05$ & 10.2 & $0.13\pm0.03$\\
\hline 
\end{tabular} 
\tablefoot{The columns are as follows: (1) source name; (2) and (3): coordinates of the emission peak position; (4) S/N ratio of the source; (5) peak intensity; (6) integrated flux density; (7) and (8) effective radius of the clump in angular and linear units. The clumps SMM~6, SMM~8, IRAS~13042, and BLOB~1 have an effective radius smaller than half the 
beam FWHM, and are hence considered as unresolved sources.\tablefoottext{a}{The clumps BLOB~1 and BLOB~2 were not identified as individual clumps of G304.74 in Paper~I.}}
\end{minipage}
\end{table*}

\begin{table*}
\caption{SABOCA 350 $\mu$m cores in G304.74.}
\begin{minipage}{2\columnwidth}
\centering
\renewcommand{\footnoterule}{}
\label{table:SABOCA}
\begin{tabular}{c c c c c c c c}
\hline\hline 
Source & $\alpha_{2000.0}$ & $\delta_{2000.0}$ & S/N & $I_{350}^{\rm peak}$ & $S_{350}$ & \multicolumn{2}{c}{$R_{\rm eff}$}\\

       & [h:m:s] & [$\degr$:$\arcmin$:$\arcsec$] & & [Jy~beam$^{-1}$] & [Jy] & [\arcsec] & [pc]\\
\hline
SMM 1a & 13 06 19.38 & -61 30 19.32 & 9.0 & $1.78\pm0.58$ & $4.04\pm1.24$ & 9.3 & $0.11\pm0.03$\\
SMM 1b & 13 06 23.36 & -61 30 10.37 & 4.5 & $1.02\pm0.37$ & $4.15\pm1.28$ & 8.1 & $0.10\pm0.03$\\
SMM 2 & 13 06 28.82 & -61 29 43.43 & 6.7 & $1.34\pm0.45$ & $2.18\pm0.69$ & 6.7 & $0.08\pm0.02$\\
SMM 3 & 13 06 37.00 & -61 28 51.00 & 10.1 & $2.02\pm0.65$ & $3.58\pm1.11$ & 8.4 & $0.10\pm0.03$ \\
SMM 4a & 13 06 46.21 & -61 28 48.04 & 8.4 & $1.67\pm0.55$ & $4.98\pm1.53$ & 10.5 & $0.13\pm0.03$\\
SMM 4b & 13 06 46.84 & -61 28 22.54 & 5.9 & $1.19\pm0.41$ & $3.24\pm1.01$ & 7.8 & $0.10\pm0.02$\\
SMM 4c & 13 06 42.86 & -61 28 33.03 & 4.6 & $0.93\pm0.35$ & $1.50\pm0.50$ & 5.0 & $0.06\pm0.02$\\
IRAS 13037-6112a & 13 06 51.45 & -61 28 24.04 & 11.0 & $2.20\pm0.70$ & $3.74\pm1.15$ & 8.3 & $0.10\pm0.03$ \\
IRAS 13037-6112b & 13 06 51.24 & -61 27 51.04 & 6.2 & $1.23\pm0.42$ & $2.27\pm0.72$ & 6.6 & $0.08\pm0.02$ \\
SMM 6a & 13 06 55.42 & -61 27 27.03 & 7.6 & $1.52\pm0.50$ & $4.97\pm0.70$ & 10.0 & $0.12\pm0.03$ \\
SMM 6b & 13 06 52.08 & -61 27 40.54 & 5.0 & $1.00\pm0.37$ & $1.00\pm0.37$ & $<4.5$ & $<0.07$ \\
SMM 7 & 13 07 04.20 & -61 26 15.00 & 5.6 & $1.13\pm0.40$ & $2.12\pm0.68$ & 6.3 & $0.08\pm0.02$\\
IRAS 13039-6108a & 13 07 06.49 & -61 24 32.98 & 11.1 & $2.22\pm0.70$ & $5.88\pm1.80$ & 11.4 & $0.14\pm0.04$\\
IRAS 13039-6108b & 13 07 08.58 & -61 23 56.97 & 5.2 & $1.05\pm0.38$ & $1.21\pm0.42$ & 4.9 & $0.06\pm0.02$\\
SMM 9 & 13 07 12.53 & -61 22 46.43 & 11.8 & $2.35\pm0.74$ & $3.61\pm1.12$ & 8.1 & $0.10\pm0.03$\\
IRAS 13042-6105 & 13 07 20.66 & -61 21 52.33 & 5.9 & $1.18\pm0.41$ & $1.27\pm0.44$ & 5.1 & $0.06\pm0.02$\\
\hline 
BLOB 2 & 13 06 39.50 & -61 30 01.51 & 4.0 & $0.81\pm0.31$ & $0.81\pm0.31$ & $<4.5$ & $<0.07$ \\
\hline 
\end{tabular} 
\tablefoot{The meaning of columns are as in Table~\ref{table:LABOCA}. The LABOCA sources SMM~5, SMM~8, and BLOB~1 were not detected in the SABOCA map.}
\end{minipage}
\end{table*}

\subsection{\textit{WISE} infrared counterparts of the submillimetre sources}

To examine which percentage of the SABOCA cores have an IR counterpart, and are hence candidates for hosting YSOs, we cross-matched our SABOCA source catalogue with the AllWISE Source Catalogue\footnote{\url{http://irsa.ipac.caltech.edu/data/download/wise-allwise/}.}, which contains $\sim 7.5 \times 10^8$ sources. As a matching radius, we used a value of $9\arcsec$, which corresponds to the SABOCA beam FWHM. For the LABOCA clumps SMM~5, SMM~8, and BLOB~1, which were not detected in our $9\arcsec$ resolution SABOCA map, we performed the cross-match with the AllWISE catalogue using the LABOCA peak position. The \textit{WISE} IR counterparts of the submm sources in G304.74, and their Vega magnitudes in the four \textit{WISE} bands are listed in Table~\ref{table:wise}. 

We found that 14 out of the 20 submm sources listed in Table~\ref{table:wise} 
have a counterpart in the AllWISE catalogue. However, the \textit{WISE} catalogue contains also other types of IR sources than YSOs, such as foreground stars and extragalactic objects, which can be seen in projection towards the submm sources. On the other hand, the fairly high latitude of G304.74 reduces 
the likelihood of contamination by background stars. 

To investigate the nature of the identified \textit{WISE} sources, we employed the \textit{WISE} colour criteria by Koenig et al. (2012; their Appendix), 
which are based on earlier \textit{Spitzer} colour criteria (e.g. \cite{gutermuth2008}). In particular, the Class~I YSO candidates are red objects whose \textit{WISE} colours are expected to obey the inequalities $[3.4]-[4.6]>1.0$ and $[4.6]-[12]>2.0$. These colour criteria are fulfilled by all the other \textit{WISE} sources except those associated with SMM~1b, SMM~6c, SMM~7, and SMM~8. The \textit{WISE} sources seen towards SMM~1b and SMM~8 have the $[3.4]-[4.6]$ and $[4.6]-[12]$ colours consistent with a shock emission feature, or a knot ($>1.0$ and $<2.0$, respectively; \cite{koenig2012}). This could also explain the large projected separations between the SABOCA peak and the \textit{WISE} source for SMM~1b ($8\farcs9$ (0.11~pc)) and SMM~8 ($7\farcs3$ (0.09~pc)). The \textit{WISE} source lying $5\farcs6$ from the dust peak of SMM~6c has the colours $[3.4]-[4.6]=2.3$ and $[4.6]-[12]=2.0$, where the latter colour places it just in between the formal classifications of a Class~I object and a shock feature. Nevertheless, SMM~6c appears to be associated with an ongoing star formation. The \textit{WISE} source seen $6\farcs0$ of SMM~7 is so weak that it is likely to be an extragalactic object (either a star-forming galaxy or an active galactic nucleus; \cite{koenig2012}). 

For comparison, we also cross-matched our SABOCA core catalogue with the \textit{WISE} YSO catalogue of Marton et al. (2016). The latter catalogue is based on the aforementioned AllWISE catalogue, which was explored by Marton et al. (2016) using the Support Vector Machine as a supervised machine learning method. This way, the authors were able to effectively remove the contaminants, such as extragalactic sources and evolved stars. The final Class~I and II YSO candidate catalogue of Marton et al. (2016) contains 133\,980 sources. However, using the same matching radius of $9\arcsec$ as above, only the \textit{WISE} source associated with SMM~2 was found in the Marton et al. (2016) catalogue. Such low counterpart matching rate can, at least partly, be explained by the initial selection made by Marton et al. (2016), that is the authors selected only the brightest \textit{WISE} sources (${\rm S/N}>3$) that benefit from high-quality near-IR photometry (errors $<0.1$).

To conclude, 13 out of the 20 submm sources listed in Table~\ref{table:wise} appear to be associated with star formation, and hence the candidate YSO percentage among the sources is $65\% \pm 18 \%$. Other way round, the percentage of IR-dark objects in G304.74 is $35\% \pm 13 \%$.

\begin{table*}
\caption{\textit{WISE} IR sources detected towards the submm sources in G304.74.}
{\small
\begin{minipage}{2\columnwidth}
\centering
\renewcommand{\footnoterule}{}
\label{table:wise}
\begin{tabular}{c c c c c c c c}
\hline\hline 
Source & \textit{WISE} ID & W1(3.4 $\mu$m) & W2(4.6 $\mu$m) & W3(12 $\mu$m) & W4(22 $\mu$m) & Offset & Type\tablefootmark{a}\\
       &  & [mag] & [mag] & [mag] & [mag] & [$\arcsec$] & \\
\hline
SMM 1a\tablefootmark{b} & \ldots & \ldots & \ldots & \ldots & \ldots & \ldots & \ldots\\ 
SMM 1b & J130622.27-613014.7 & $10.511 \pm 0.024$ & $8.712 \pm0.02$ & $6.845 \pm0.045$ & $4.197\pm0.045$ & 8.9 & shock \\ 
SMM 2 & J130628.64-612942.7 & $9.966 \pm0.023$ & $8.489\pm 0.020$ & $5.732\pm0.028$ & $3.310\pm0.029$ & 1.4 & YSO\\
SMM 3 & J130637.23-612848.8 & $>15.531$ & $11.005\pm0.033$ & $7.667 \pm 0.087$ & $3.584 \pm 0.035$ & 2.7 & YSO \\
SMM 4a & J130646.35-612855.3 & $14.719\pm0.217$ & $12.666 \pm0.026$ & $>9.344$ & $6.001 \pm 0.233$ & 7.4 & YSO\\
SMM 4b & J130647.85-612819.5 & $11.546 \pm 0.032$ & $9.862 \pm0.026$ & $5.347 \pm 0.026$ & $1.869 \pm 0.029$ & 7.9 & YSO\\
SMM 4c\tablefootmark{b} & \ldots & \ldots & \ldots & \ldots & \ldots & \ldots & \ldots\\ 
SMM 5\tablefootmark{b,c} & \ldots & \ldots & \ldots & \ldots & \ldots & \ldots & \ldots\\ 
IRAS 13037-6112a & J130651.93-612822.2 & $8.613\pm0.025$ & $6.847\pm0.021$ & $4.030\pm0.015$ & $0.895\pm0.017$ & 3.9 & YSO\\
IRAS 13037-6112b & J130651.58-612751.7 & $13.551\pm0.108$ & $9.600\pm0.023$ & $6.168\pm0.029$ & $1.847\pm0.012$ & 2.6 & YSO \\
SMM 6a & J130654.93-612725.9 & $13.435\pm0.086$ & $10.075\pm0.023$ & $6.347\pm0.024$ & $3.570\pm0.026$ & 3.7 & YSO\\
SMM 6b & J130652.20-612735.0 & $11.896\pm0.030$ & $9.589\pm0.021$ & $7.596\pm0.049$ & $3.385\pm0.021$ & 5.6 & YSO/shock\\
SMM 7 & J130704.50-612620.6 & $>16.444$ & $>16.831$ & $>10.699$ & $5.765 \pm 0.041$ & 6.0 & SFG/AGN\\
IRAS 13039-6108a & J130707.22-612438.5 & $10.261 \pm0.032$ & $8.256 \pm0.020$ & $3.799 \pm0.014$ & $0.875 \pm0.017$ & 7.7 & YSO \\
IRAS 13039-6108b\tablefootmark{b} & \ldots & \ldots & \ldots & \ldots & \ldots & \ldots & \ldots\\ 
SMM 8\tablefootmark{c} & J130708.93-612346.9 & $13.415\pm0.084$ & $11.215\pm0.023$ & $>9.752$ & $6.015\pm0.062$ & 7.3 & shock \\ 
SMM 9\tablefootmark{b} & \ldots & \ldots & \ldots & \ldots & \ldots & \ldots & \ldots\\ 
IRAS 13042-6105 & J130721.38-612149.9 & $12.440 \pm0.041$ & $8.855 \pm0.020$ & $6.265 \pm0.027$ & $3.025 \pm0.019$ & 5.7 & YSO\\
\hline 
BLOB 1\tablefootmark{c} & J130641.48-612916.3 & $14.727\pm0.126$ & $12.233\pm0.032$ & $>9.792$ & $>8.719$ & 8.6 & YSO\\
BLOB 2\tablefootmark{b} & \ldots & \ldots & \ldots & \ldots & \ldots & \ldots & \ldots \\  
\hline 
\end{tabular} 
\tablefoot{The magnitudes are given in the Vega system, and they were taken from the AllWISE catalogue. The catalogue magnitudes were measured with profile-fitting photometry, and they represent the total in-band brightnesses. The magnitude upper limits refer to the $95\%$ confidence brightness limit.\tablefoottext{a}{Classification of the \textit{WISE} source on the basis of the Koenig et al. (2012) colour criteria. All the other sources are expected to be embedded YSOs, except that SMM~1b and SMM~8 are associated with shock emission knots, SMM~6b has the colours in between an embedded YSO and a shock feature, and the \textit{WISE} source seen towards SMM~7 is likely to be an extragalactic object, that is a star-forming galaxy or an active galactic nucleus.}\tablefoottext{b}{No \textit{WISE} counterpart was found within $9\arcsec$. We note that there is a \textit{WISE} source $9\farcs1$ away from SMM~9 (J130713.09-612238.2; ${\rm W1}=12.765 \pm 0.051$, ${\rm W2}=11.929 \pm 0.029$, ${\rm W3}>10.159$, ${\rm W4}=7.549 \pm 0.117$). Its \textit{WISE} colours are not consistent with an embedded YSO, but it could be a foreground star (\cite{koenig2012}).}\tablefoottext{c}{The source was not detected in the SABOCA map, and hence the \textit{WISE} counterpart refers to the LABOCA peak position.}}
\end{minipage}}
\end{table*}

\subsection{Physical properties of the LABOCA and SABOCA sources}

In this subsection, we revise the basic physical properties of the LABOCA clumps from 
those presented in Papers~I and II, and derive the physical properties for the SABOCA 
cores identified in the present work.  

First, we derived the dust temperatures ($T_{\rm dust}$) for the LABOCA clumps using the \textit{Herschel}/SPIRE flux density ratios (see e.g. Eqs.~(4a) and (4b) in \cite{shetty2009}). The advantage of using ratios among flux densities at three different wavelengths (250, 350, and 500~$\mu$m in our case) is that no assumption about the value of the dust emissivity index $\beta$ is required (except that the value is assumed to be constant across the source). For the purpose of this calculation, the SPIRE 250~$\mu$m and 350~$\mu$m images were smoothed to the resolution of the SPIRE 500~$\mu$m image ($36\farcs3$), and also regridded to the pixel size of the 500~$\mu$m image ($14\arcsec$). As the source flux densities we used the peak surface brightness values owing to the large beam size of the maps. The derived temperatures are tabulated in Table~\ref{table:laboca_properties}. Although the \textit{Herschel} images used here have not been calibrated to an absolute intensity scale, the employed method of using flux density ratios in the calculation of $T_{\rm dust}$ is not (significantly) subject to the absolute flux density values. However, construction of a $T_{\rm dust}$ map derived using a pixel-by-pixel modified black-body (MBB) fitting would require the properly calibrated data. We also note that a value of $T_{\rm dust}=22.2$~K derived for IRAS~13039 is very similar to a dust temperature of 22~K derived 
in Paper~I through fitting the source spectral energy distribution (SED) 
with a MBB function, and comparable to the C$^{17}$O rotation temperature of $T_{\rm rot}=18$~K derived by Fontani et al. (2005). The MBB-inferred dust temperature of 22~K for IRAS~13037 in Paper~I is also comparable to the present estimate of 20~K.

The total (gas plus dust) masses of the LABOCA and SABOCA sources were estimated from the integrated flux densities using 
the standard optically thin dust emission formulation and the dust temperatures derived above (see e.g. Eq.~(6) in Paper~I). 
The dust opacities per unit dust mass at 870~$\mu$m and 350~$\mu$m were assumed to be $\kappa_{{\rm 870\, \mu m}}=1.38$~cm$^2$~g$^{-1}$ and $\kappa_{{\rm 350\, \mu m}}=7.84$~cm$^2$~g$^{-1}$. These values are based on the Ossenkopf \& Henning (1994, hereafter OH94) model, which describes graphite-silicate dust grains that have coagulated and accreted thin ice mantles over a period of $10^5$~yr at a gas density of $n_{\rm H}= n({\rm H}) + 2 \times n({\rm H_2})\simeq 2\times n({\rm H_2}) = 10^5$~cm$^{-3}$. For comparison, in Papers~I and II we assumed that the OH94 dust grains have thick ice mantles, and used an 870~$\mu$m opacity of $\kappa_{{\rm 870\, \mu m}}\simeq1.7$~cm$^2$~g$^{-1}$. However, the present assumption of grains with thin ice mantles is supported by the finding of no significant CO depletion in Paper~II. The dust-to-gas mass ratio was fixed at a canonical value of $R_{\rm dg}=1/100$. The mass uncertainties were propagated from the flux density uncertainty and kinematic distance uncertainty ($M \propto S d^2$). The derived masses of the LABOCA clumps are listed in column~(3) in Table~\ref{table:laboca_properties}, while those of the SABOCA cores are tabulated in column~(2) in Table~\ref{table:saboca_properties}. For all the SABOCA cores that belong to a common clump (e.g. SMM~1a and SMM~1b), the dust temperature was assumed to be that of the parent clump (e.g. 14~K for both SMM~1a and SMM~1b).

The beam-averaged peak H$_2$ column densities, $N({\rm H_2})$, were derived from the peak intensities of dust emission following Eq.~(8) in Paper~I. All the assumptions were the same as above in the mass calculation. The mean molecular weight per H$_2$ molecule, which is needed in the calculation of $N({\rm H_2})$, was assumed to be $\mu_{\rm H_2}=2.82$, which is based on the assumption that the He/H abundance ratio is 0.1 (\cite{kauffmann2008}). The derived values are listed in column~(4) in Table~\ref{table:laboca_properties}, and column~(3) in Table~\ref{table:saboca_properties}. The uncertainties in $N({\rm H_2})$ were propagated from the intensity uncertainties. 

Finally, we derived the volume-averaged H$_2$ number densities, $\langle n({\rm H_2}) \rangle$, for the LABOCA clumps and SABOCA cores. For this calculation, we used Eq.~(7) of Paper I, which is based on the assumption of a spherical source with radius $R_{\rm eff}$. The values are tabulated in the last columns of Tables~\ref{table:laboca_properties} and \ref{table:saboca_properties}, where the quoted uncertainties were propagated from those of $M$ and $R_{\rm eff}$.

Regarding the statistics of the derived physical properties, the mean value of the dust temperature was found to be $\langle T_{\rm dust} \rangle = 15.0 \pm 0.8$~K, where the quoted uncertainty is the standard error of the mean (i.e. $\sigma/\sqrt{N}$, where $\sigma$ is the standard deviation, and $N$ the sample size). The mean values of the mass, H$_2$ column density, and H$_2$ number density for the LABOCA (SABOCA) sources were found to be $55\pm10$~M$_{\sun}$ ($29\pm3$~M$_{\sun}$), $(2.0\pm0.2)\times10^{22}$~cm$^{-2}$ ($(2.9\pm0.3)\times10^{22}$~cm$^{-2}$), and $(3.1\pm0.2)\times10^4$~cm$^{-3}$ ($(7.9\pm1.2)\times10^4$~cm$^{-3}$), respectively. To calculate the sample average of the H$_2$ number densities, we applied survival analysis to take the lower limit values (right-censored data) into account. It was assumed that the right-censored data follow the same distribution as the uncensored values, and we used the Kaplan-Meier method to construct a model of the input data. For this purpose, we used the Nondetects And Data Analysis for environmental data (NADA; \cite{helsel2005}) package for {\tt R}.

\begin{table}
\caption{Physical properties of the LABOCA 870~$\mu$m clumps.}
{\small
\begin{minipage}{1\columnwidth}
\centering
\renewcommand{\footnoterule}{}
\label{table:laboca_properties}
\begin{tabular}{c c c c c}
\hline\hline 
Source & $T_{\rm dust}$ & $M$ & $N({\rm H_2})$ & $\langle n({\rm H_2}) \rangle$ \\  
       & [K] & [M$_{\sun}$] & [$10^{22}$ cm$^{-2}$] & [$10^4$ cm$^{-3}$]\\ 
\hline
SMM 1 & 14.0 & $128\pm68$ & $2.6\pm0.4$ & $1.5\pm0.8$\\
SMM 2 & 12.5 & $64\pm34$ & $2.4\pm0.3$ & $2.4\pm1.3$\\
SMM 3 & 13.7 & $71\pm38$ & $2.5\pm0.4$ & $3.6\pm1.9$\\
SMM 4 & 13.2 & $136\pm73$ & $3.5\pm0.5$ & $3.8\pm2.0$\\
IRAS 13037 & 20.0 & $59\pm31$ & $1.4\pm0.2$ & $3.5\pm1.8$\\
SMM 5 & 14.0 & $23\pm13$ & $1.4\pm0.3$ & $2.9\pm1.6$\\
SMM 6 & 14.1 & $34\pm18$ & $2.2\pm0.3$ & $>1.5$\tablefootmark{a}\\
SMM 7 & 14.5 & $41\pm22$ & $1.7\pm0.3$ & $2.9\pm1.5$\\
IRAS 13039 & 22.2 & $58\pm31$ & $1.4\pm0.2$ & $2.1\pm1.1$\\
SMM 8 & 13.9 & $24\pm13$ & $1.6\pm0.3$ & $>1.1$\tablefootmark{a}\\
SMM 9 & 16.3 & $54\pm29$ & $1.9\pm0.3$ & $4.5\pm2.4$\\
IRAS 13042 & 15.1 & $17\pm10$ & $1.1\pm0.2$ & $>0.7$\tablefootmark{a}\\
BLOB 1 & 11.5 & $39\pm21$ & $2.6\pm0.4$ & $>1.8$\\
BLOB 2 & 14.9 & $23\pm13$ & $1.3\pm0.2$ & $3.6\pm2.0$\\
\hline 
\end{tabular} 
\tablefoot{\tablefoottext{a}{The effective radius of the clump was unresolved, and hence only a lower limit to the number density could be calculated. The uncertainty in the density is taken into account in the quoted lower limit value.}}
\end{minipage}  }
\end{table}

\begin{table}
\caption{Physical properties of the SABOCA 350~$\mu$m cores.}
{\small
\begin{minipage}{1\columnwidth}
\centering
\renewcommand{\footnoterule}{}
\label{table:saboca_properties}
\begin{tabular}{c c c c}
\hline\hline 
Source & $M$ & $N({\rm H_2})$ & $\langle n({\rm H_2}) \rangle$ \\  
       & [M$_{\sun}$] & [$10^{22}$~cm$^{-2}$] & [$10^4$~cm$^{-3}$]\\ 
\hline
SMM 1a & $31\pm18$ & $4.3\pm1.4$ & $7.9\pm4.8$\\
SMM 1b & $31\pm19$ & $2.5\pm0.9$ & $10.8\pm6.5$\\
SMM 2 & $24\pm15$ & $4.7\pm1.6$ & $16.0\pm9.7$\\
SMM 3 &  $29\pm18$ & $5.2\pm1.7$ & $10.0\pm6.0$\\
SMM 4a &  $46\pm28$ & $4.9\pm1.6$ & $7.1\pm4.3$\\
SMM 4b &  $30\pm18$ & $3.5\pm1.2$ & $10.1\pm6.1$\\
SMM 4c &  $14\pm8$ & $2.7\pm1.0$ & $21.8\pm13.4$\\
IRAS 13037a & $11\pm7$ & $2.0\pm0.6$ & $3.7\pm2.2$\\
IRAS 13037b & $7\pm4$ & $1.1\pm0.5$ & $4.4\pm2.7$\\
SMM 6a & $37\pm20$ & $3.6\pm1.2$ & $7.3\pm4.0$\\
SMM 6b & $7\pm5$ & $2.4\pm0.9$ & $>2.7$\tablefootmark{a}\\
SMM 7 & $14\pm9$ & $2.4\pm0.9$ & $9.7\pm5.9$\\
IRAS 13039a & $13\pm8$ & $1.6\pm0.5$ & $1.7\pm1.0$\\
IRAS 13039b & $3\pm2$ & $7.6\pm0.3$ & $4.4\pm2.7$\\
SMM 9 & $18\pm11$ & $3.6\pm1.1$ & $6.0\pm3.6$\\
IRAS 13042 & $8\pm5$ & $2.3\pm0.8$ & $12.2\pm7.6$\\
BLOB 2 & $5\pm3$ & $1.6\pm0.6$ & $>1.8$\tablefootmark{a}\\
\hline 
\end{tabular} 
\tablefoot{The quoted physical properties of the cores were calculated by assuming the parent clump dust temperatures from column (2)~in Table~\ref{table:laboca_properties}.\tablefoottext{a}{The effective radius of the core was unresolved, and hence only a lower limit to the number density could be calculated. The uncertainty in the density is taken into account in the quoted lower limit value.}}
\end{minipage}  }
\end{table}

\section{Discussion}

\subsection{General characteristics of the filamentary structure of G304.74+01.32}

The mean dust temperature we derived towards the clumps in G304.74, $15.0\pm0.8$~K, is very similar to the average NH$_3$ rotation temperature of $T_{\rm rot}=15$~K derived for a large sample of 109 IRDCs by Chira et al. (2013). Indeed, the high gas density of the G304.74 filament (the average H$_2$ number density estimated for the LABOCA clumps is $(3.1\pm0.2)\times10^4$~cm$^{-3}$) has likely led to the coupling of the gas and dust temperatures (\cite{goldsmith2001}).  

The total projected length of G304.74, as seen in both the LABOCA and SABOCA maps, is $\sim12\arcmin$, or $\sim 8.9\pm2.3$~pc. On the other hand, the mean effective radius of the LABOCA (SABOCA) filament is $14\farcs1$ ($7\farcs4$). Hence, the aspect ratio of the LABOCA (SABOCA) filament is $r_{\rm aspect}\simeq26$ ($r_{\rm aspect}\simeq49$). The projected width of the LABOCA filament in linear units is $0.35\pm0.09$~pc, while that of the SABOCA filament is $0.18\pm0.05$~pc. The latter value is consistent with a width of $\sim0.1$~pc, which is found to be fairly common among interstellar filaments, and it is also comparable with the width of a filamentary cloud expected in the scenario of filament formation as a result of converging turbulent flows (e.g. \cite{palmeirim2013}, and references therein). Overall, the centrally concentrated dust distribution of G304.74 as seen in the SABOCA map resembles the SABOCA maps of IRDCs by Ragan et al. (2013). 

The total mass of the LABOCA filament can be estimated to be 
$773\pm131$~M$_{\sun}$ as the sum of the clump masses from 
Table~\ref{table:laboca_properties}. Similarly, the sum of the SABOCA core masses (Table~\ref{table:saboca_properties}) suggests that the SABOCA filament's mass is $327\pm56$~M$_{\sun}$, that is about 42\% of the mass of the LABOCA filament. The line masses of the LABOCA and SABOCA filaments are hence estimated to be $M_{\rm line}^{\rm LABOCA}=87\pm27$~M$_{\sun}$~pc$^{-1}$ and $M_{\rm line}^{\rm SABOCA}=37\pm11$~M$_{\sun}$~pc$^{-1}$. The former value is smaller by a factor of $1.55 \pm 0.48$ than the line mass estimated in Paper~II owing to the different method how the source flux densities were determined and the different assumptions about the dust temperature. 

For an unmagnetised, infinite, isothermal filament, the instability is
reached if the $M_{\rm line}$ value exceeds the critical equilibrium value of
$M_{\rm line}^{\rm crit}=2c_{\rm s}^2/G$, where $c_{\rm s}$ is the sound speed and 
$G$ the gravitational constant (e.g. \cite{ostriker1964}; 
\cite{inutsuka1992}). If the gas kinetic temperature is comparable to the 
average dust temperature of the filament, $\sim15$~K, the critical line mass 
is $\sim25$~M$_{\sun}$~pc$^{-1}$. Hence, the LABOCA and SABOCA filaments would 
be thermally supercritical by factors of $\sim 3.5 \pm 1.1$ and $\sim 1.5 \pm 0.4$. The latter values should be considered as lower limits owing to the emission that is being filtered out in our submm maps (Sect.~3.1). We note that if the non-thermal (turbulent) motions are important on the larger scale of LABOCA emission, the thermal sound speed in the aforementioned $M_{\rm line}^{\rm crit}$ formula should be replaced by the total velocity dispersion (see Paper~II and references therein). In Paper~II, we estimated that the total (thermal plus non-thermal) C$^{17}$O line velocity dispersion in G304.74 is, on average, $\sigma_{\rm v}^{\rm tot}=0.54$~km~s$^{-1}$ (under the assumption that $T_{\rm gas}=15$~K). This velocity dispersion would lead to a critical line mass of $\sim136$~M$_{\sun}$~pc$^{-1}$. In this case, the LABOCA filament would have a $M_{\rm line}/M_{\rm line}^{\rm crit}$ ratio of $0.64 \pm 0.20$, that is the cloud would be marginally subcritical, but near virial equilibrium within the uncertainties (see Paper~II). Again, considering the filtered-out low-column density dust emission in the LABOCA map (Sect.~3.1), which leads to an underestimated mass (and hence $M_{\rm line}$), a near virial equilibrium of the G304.74 LABOCA filament appears more likely. Moreover, non-thermal motions are unlikely to provide uniform support against gravity along the whole filament, in which case the G304.74 LABOCA filament could be supercritical. On the other hand, the width of the SABOCA filament ($0.18\pm0.05$~pc) is comparable to the sonic scale below which supersonic turbulence becomes subsonic (e.g. \cite{vazquez2003}; \cite{palmeirim2013}). This supports the usage of the thermal sound speed in the calculation of $M_{\rm line}^{\rm crit}$ above.

We conclude that the G304.74 filament is (marginally) supercritical, because the cloud is fragmented into clumps and cores, and  most of the cores are hosting embedded YSOs. Indeed, hierarchical fragmentation and the presence of gravitationally collapsed cores hosting newly forming stars are both expected to be found in a thermally supercritical filament, which is susceptible to fragmentation.

A potentially important role in the dynamics of filamentary molecular clouds is that played by magnetic fields. The dynamical analysis presented In Paper~II suggests that the overall effect of the magnetic field on G304.74 is to provide support, and that the magnetic field configuration is poloidally dominated (see \cite{fiege2000a}). Similarly, the Snake IRDC G11.11-0.12 is likely dominated by the poloidal component of the magnetic field, and hence magnetically supported (\cite{fiege2004}). Fiege \& Pudritz (2000b) found that model filaments with purely toroidal magnetic fields are stable against sausage instability modes, while model clouds with a poloidal field component can be unstable. This is consistent with a scenario where G304.74 fragmented into clumps as a result of sausage-type instability (see Paper~II). The upcoming \textit{Herschel} GBS data sets for the Coalsack region\footnote{\url{http://www.herschel.fr/cea/gouldbelt/en/Phocea/Vie_des_labos/Ast/ast_visu.php?id_ast=66}.} allow us to map the H$_2$ column density distribution of G304.74, and to derive the filament's density profiles. The density distributions, in turn, can be used to constrain the magnetic model that best describes the G304.74 filament (e.g. \cite{johnstone2003}).

The \textit{Herschel}/SPIRE images shown in Fig.~\ref{figure:spire} reveal the presence of dusty striations on the western side of G304.74. The most prominent of these striations, which we call striation~1 and striation~2, appear to be oriented perpendicular to the G304.74 filament. Moreover, some of the striations, most notably striation~1, can be seen in absorption in the \textit{WISE} 12~$\mu$m image shown in Fig.~\ref{figure:wise}. The G304.74 
filament could be in the process of accreting material along the striations, similarly to what has been suggested for the Taurus B211 filament (\cite{palmeirim2013}). Depending on the kinematic properties of such accretion flows, they could generate accretion shocks at the border of the G304.74 filament. Also, the kinetic energy of the incoming accretion flows could potentially be (partly) converted into turbulent motions, and hence contribute to the energy balance of G304.74. 

In Paper~II, we observed double-peaked $^{13}$CO$(2-1)$ lines towards the southern part of the G304.74 filament (towards the LABOCA clumps SMM~1--4 and IRAS~13037), which is an indication of gas infall motions. The infall motions, in turn, could be related to the aforementioned \textit{Herschel}/SPIRE striations. The infall velocities estimated in Paper~II range from ${\rm v}_{\rm inf}\sim0.03$~km~s$^{-1}$ to 0.2~km~s$^{-1}$ (0.12~km~s$^{-1}$ on average), which is suggestive of subsonic gas flows that do not create shocks. The corresponding mass infall rates were estimated to be $\dot{M}_{\rm inf}\sim 2-36 \times 10^{-5}$~M$_{\sun}$~yr$^{-1}$ with a mean value of $\sim 1.5 \times 10^{-4}$~M$_{\sun}$~yr$^{-1}$. However, higher resolution spectral line imaging would be needed to examine the kinematics and gas dynamics of G304.74 and its surroundings further.

If the G304.74 filament has settled into near virial equilibrium as argued above, one would expect that the age of the filament is comparable to its signal crossing time, that is $\tau_{\rm fil}\sim\tau_{\rm cross}=2 \times R_{\rm fil}/\sigma_{\rm fil}$, where $R_{\rm fil}$ is the radius of the filament, and $\sigma_{\rm fil}$ the gas velocity dispersion (e.g. \cite{hernandez2012}). If $\sigma_{\rm fil}$ is 0.54~km~s$^{-1}$ as we estimated in Paper~II, the crossing time for the G304.74 LABOCA filament would be $\sim 6\times10^5$~yr. On the other hand, one potential scenario for the formation of filamentary IRDCs (and other types of filamentary molecular clouds) is that they originate in the merging process of converging gas flows (e.g. \cite{jimenez2010}; \cite{beuther2014}, and references therein). Following the analysis of Hernandez et al. (2012; Sect.~5 therein), the time required for the G304.74 LABOCA filament to form through the merging of two converging flows would be $\tau_{\rm form}\sim1.2$~Myr, but this estimate should be taken as a rough estimate only because of the unknown velocities and gas densities of the colliding parent flows. Nevertheless, the aforementioned timescale estimates suggest that G304.74 could have reached virial equilibrium even in the case the filament was formed only about 1.2~Myr ago. A similar conclusion was reached by Hernandez et al. (2012) for the filamentary IRDC~G035.39-00.33. 

The \textit{WISE} colours of the YSOs found in G304.74 refer to Class~I objects (Sect.~3.2). The lifetime of such objects is estimated to be $\sim0.5$~Myr (\cite{evans2009}), and less for higher mass YSOs. Hence, the formation of stars in G304.74 might have started soon after the cloud was formed, or already during the cloud formation process (cf. the aforementioned crossing time of $\tau_{\rm cross}\sim 0.6$~Myr). One potential route for a rapid star formation is the elevated protostellar accretion rates at the early cloud formation stages owing to the cloud-forming supersonic colliding flows (e.g. \cite{banerjee2009}).

\begin{figure*}[!htb]
\begin{center}
\includegraphics[scale=0.5]{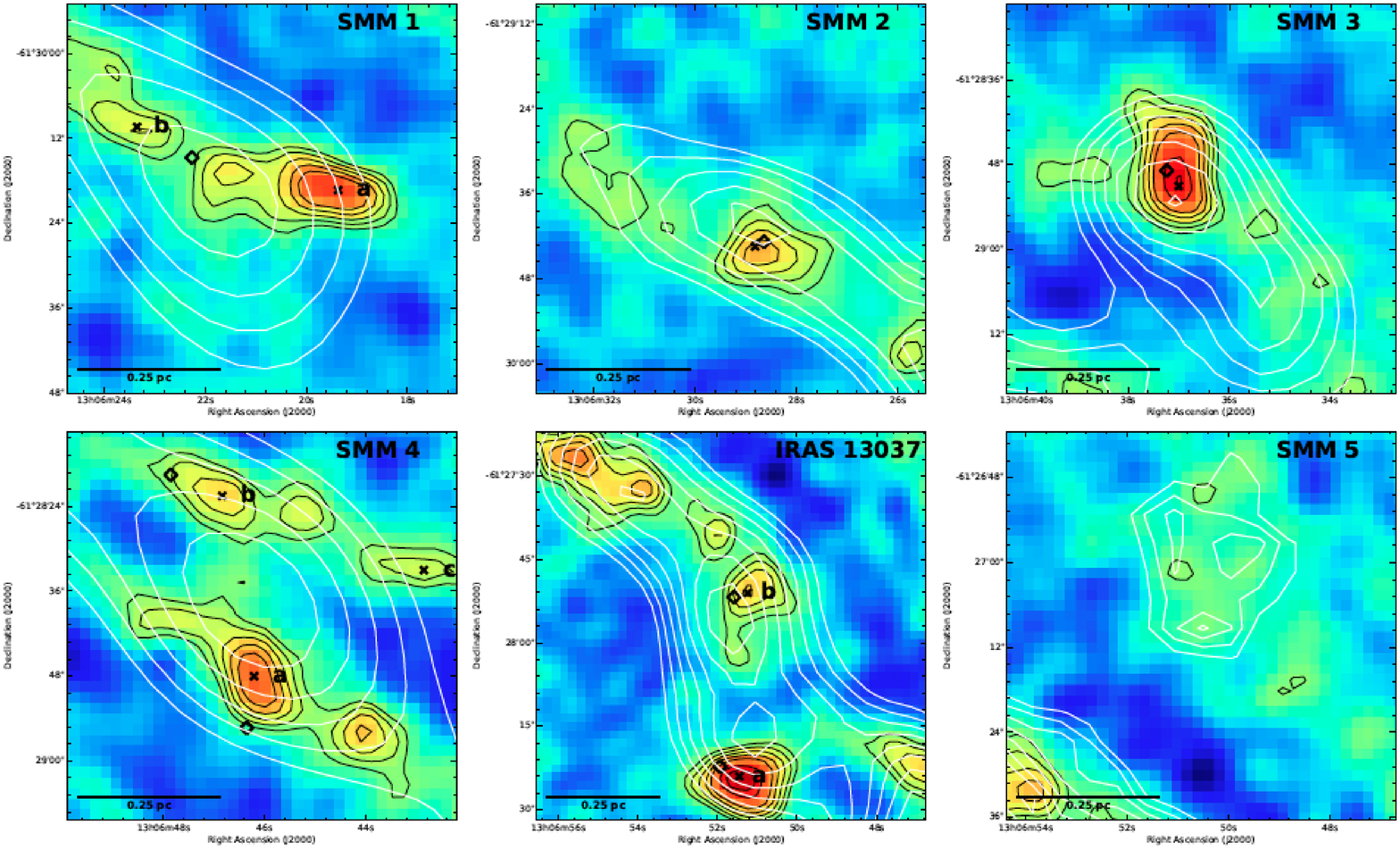}
\includegraphics[scale=0.5]{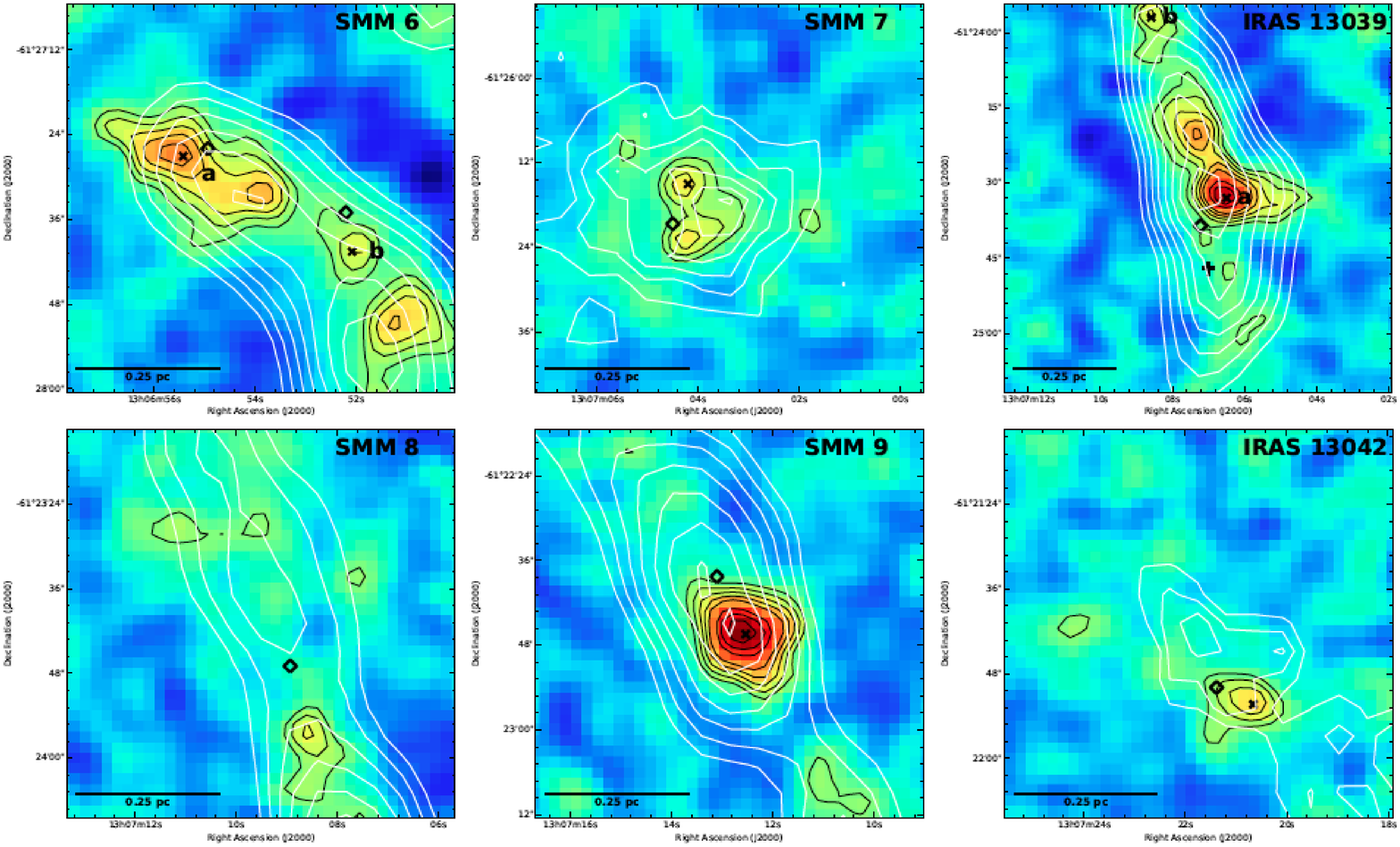}
\caption{Zoom-in images towards the clumps in G304.74. In each panel, the white contours represent the LABOCA 870~$\mu$m emission, while the colour scale (linear scaling) and black contours show the SABOCA 350~$\mu$m emission. The contour levels are as in Fig.~\ref{figure:submm}. Each image is centred on the LABOCA peak position of the clump, and is $55\arcsec \times 55\arcsec$ in size with the exception of the IRAS~13037 and IRAS~13039 images, which are $70\arcsec$ and $77\farcs4$ on a side, respectively. The crosses indicate the peak positions of the SABOCA 350~$\mu$m cores, while the diamond symbols show the positions of the \textit{WISE} sources from Table~\ref{table:wise}. The plus sign in the IRAS~13039 image indicates the position of the 18~GHz and 22.8~GHz radio continuum source found by S{\'a}nchez-Monge et al. (2013). A scale bar of 0.25~pc is shown in the bottom left corner of each panel.}
\label{figure:fragments}
\end{center}
\end{figure*}

\addtocounter{figure}{-1}
\begin{figure*}
\begin{center}
\includegraphics[scale=0.5]{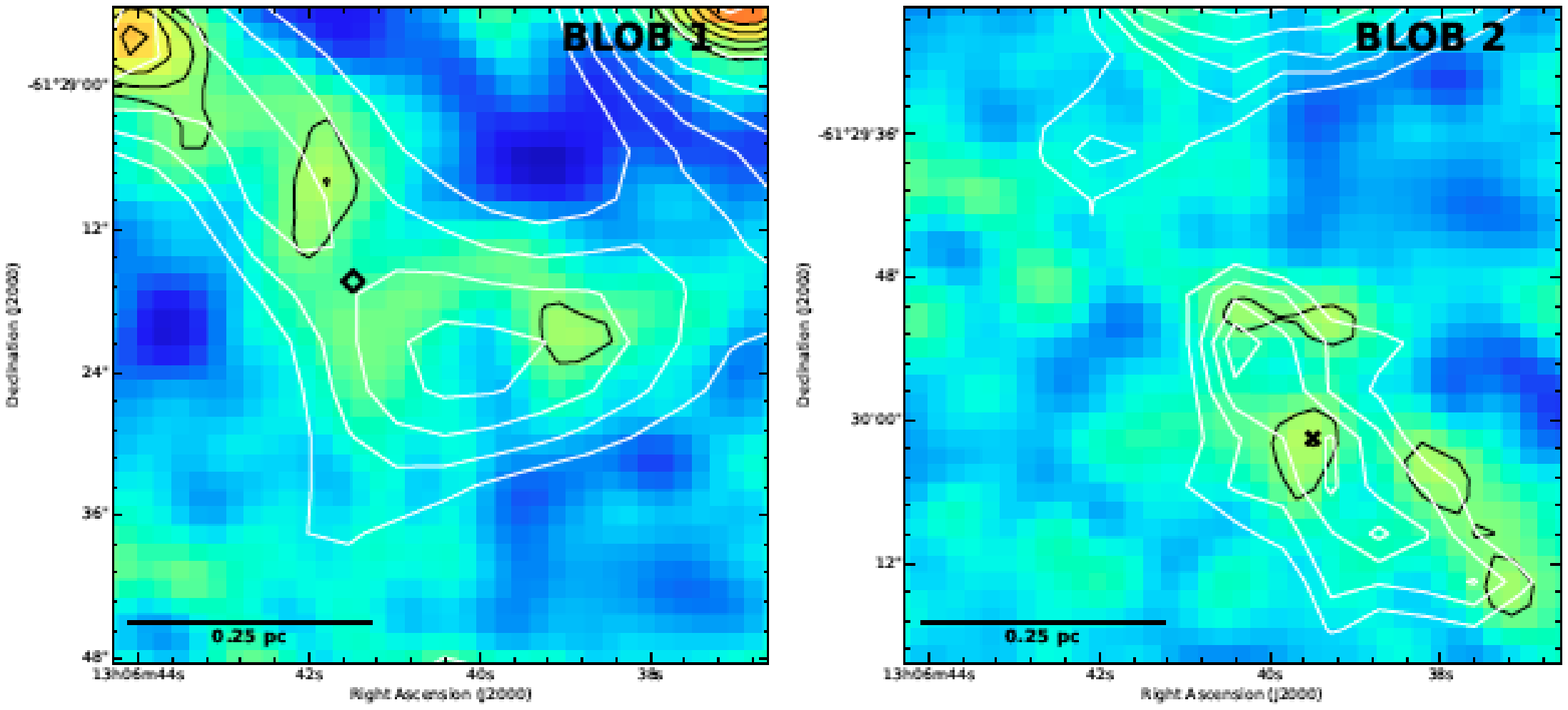}
\caption{continued.}
\label{figure:fragments}
\end{center}
\end{figure*}

\subsection{Hierarchical fragmentation of the G304.74+01.32 filament}

In Paper~II, we suggested that the G304.74 LABOCA filament has fragmented 
into clumps through a sausage-type fluid instability mechanism. The chief argument behind this conclusion 
was that the observed mean projected separation between the clumps ($\sim0.75$~pc) 
is comparable with a theoretical prediction of the sausage instability framework. In the present study, 
the separations between the LABOCA clumps are found to range from 0.39~pc to 1.33~pc with a mean of $0.77\pm0.09$~pc. Although the average width of the LABOCA filament, as calculated from the clump effective radii, is only $0.35 \pm 0.09$~pc (Sect.~4.1), its average width across the $3\sigma$ emission is about 0.72~pc (based on separate measurements for the northern and southern filaments using rectangles that just enclose the $3\sigma$ contours). The observed clump separation interval is fairly similar to the aforementioned total width of the filament, which is also expected from the fragmentation of a magnetised, cylindrical cloud (\cite{nakamura1993}).

In the case of gravitational fragmentation of a finite, 
cylindrical cloud, the number of fragments that form along the cylinder can be 
expressed as $N_{\rm frag}\simeq0.5 \times r_{\rm aspect}$ (e.g. \cite{bastien1991}; \cite{wiseman1998}). 
If the fragmentation of the LABOCA filament into clumps were due to such gravitational instability, one 
would expect the number of clumps to be about 13 (Sect.~4.1), or $0.5 \times8.9/0.72\simeq 6$ if the aspect ratio is calculated using the aforementioned width of the $3\sigma$ emission level of the cloud. Interestingly, the former value is very close to the number of clumps (14) identified in the present work. For the denser SABOCA filament, the number of fragments expected from its aspect ratio is about 25, while the number of SABOCA cores is 17. However, because three of the LABOCA clumps were resolved out in our SABOCA map, the true number of cores is about 20 or more (see below), and hence comparable to the predicted value. An obvious caveat in this analysis is that G304.74 is not perfectly cylindrical in shape, and its true length (the longest axis) could be longer owing to a possible inclination angle of the cloud, in which case the true aspect ratio (and hence $N_{\rm frag}$) would be higher. Nevertheless, it seems possible that the densest portion of the G304.74 filament, particularly that traced by SABOCA, underwent Jeans-type gravitational fragmentation. 

The present SABOCA data show that the LABOCA clumps in G304.74 are composed of smaller cores. The zoom-in images towards all the LABOCA clumps are shown in Fig.~\ref{figure:fragments}. The percentage of clumps that are found to be fragmented into cores, or the multiplicity fraction, is $36\%\pm16\%$. This percentage should be taken as a lower limit because three of the LABOCA clumps were resolved out in our SABOCA image, and those clumps could contain two or more cores. Figure~\ref{figure:fragments} also shows that some of the cores that were treated as a single source in our {\tt BLOBCAT} analysis appear to host an additional component. These sources are SMM~1a, SMM~4a, SMM~4b, SMM~6a, SMM~7, and IRAS~13039a. If SMM~7 is indeed fragmented as suggested by Fig.~\ref{figure:fragments}, the aforementioned multiplicity fraction would be $43\%\pm17\%$. The mean core separations in the fragmented LABOCA clumps are tabulated in Table~\ref{table:jeans}. These values were calculated by taking the aforementioned additional substructures into account. On average, the cores are projectively separated by 0.22~pc. 

The observed core separations can be compared with the Jeans length

\begin{equation}
\lambda_{\rm J}=\sqrt{\frac{\pi c_{\rm s}^2}{G\langle \rho \rangle}}\,,
\end{equation}
where $\langle \rho \rangle$ is the mean mass density. The values of $\lambda_{\rm J}$ calculated using the dust temperatures and densities from Table~\ref{table:laboca_properties} are tabulated in Table~\ref{table:jeans}. The quoted uncertainties were propagated from the density uncertainties. The derived $\lambda_{\rm J}$ values are fairly similar to the observed core separations, particularly in the case of SMM~1, SMM~6, and IRAS~13039. For SMM~4 and IRAS~13037, the observed core spacings exceed the local Jeans length (by factors of $1.92\pm0.58$ and $2.56\pm0.64$), which could be an indication that the true gas temperature is higher than used in the calculation. Indeed, for SMM 4 the estimated dust temperature is only 13.2 K, but the presence of (at least) two YSOs in the clump suggests that the temperature is likely higher. 

An additional method to test whether a clump might have fragmented into cores via Jeans instability is to calculate its local thermal Jeans mass, which is defined by

\begin{equation}
M_{\rm J}=\frac{4}{3}\pi\left(\frac{\lambda_{\rm J}}{2}\right)^3\langle \rho \rangle\,.
\end{equation}
The clump's mass can be compared with its Jeans mass, and to calculate the so-called Jeans number, $N_{\rm J}=M/M_{\rm J}$. The values of $M_{\rm J}$ and $N_{\rm J}$ are listed in Table~\ref{table:jeans}. The uncertainties are large, but the Jeans numbers for IRAS~13037, SMM~6, SMM~7, and IRAS~13039 are comparable to the observed number of subcomponents (ranging from $N_{\rm J}>1$ for SMM~6 to $N_{\rm J}=5\pm3$ for IRAS~13037 and SMM~7). For SMM~1 and SMM~4, the Jeans number is considerably larger, namely $N_{\rm J}=15\pm8$ and $N_{\rm J}=22\pm13$, respectively. The true number of cores within the clumps could well be higher, and this possibility could be tested using imaging observations at higher resolution. Also, if non-thermal motions contribute to the pressure support on clump scales, the Jeans mass would be higher, and hence the Jeans number would be lower.

Overall, our Jeans analysis supports the possibility that the clumps of 
G304.74 fragmented into cores via Jeans gravitational instability. We conclude that G304.74 is fragmented in a hierarchical fashion, similar to that found for the Snake Nebula (\cite{wang2014}; \cite{ragan2015}; see also \cite{henshaw2016} for the case of the IRDC G035.39-00.33). The parent filament might have formed through the collision of interstellar turbulent flows, and then fragmented into a chain of clump-scale structures owing to gravitational and magnetic perturbations mediated by sausage-type instability. The clumps represent the density enhancements of the filament, and the physical properties of the clumps allow the Jeans gravitational instability to become the dominant fragmentation mechanism.

\begin{table*}
\caption{Projected core separations, and the Jeans instability parameters in the fragmented clumps of G304.74.}
{\normalsize
\begin{minipage}{2\columnwidth}
\centering
\renewcommand{\footnoterule}{}
\label{table:jeans}
\begin{tabular}{c c c c c c c c c}
\hline\hline 
Source & $M$\tablefootmark{a} & $N_{\rm cores}$\tablefootmark{b} & $\langle \lambda_{\rm sep}\rangle$\tablefootmark{c} & $\lambda_{\rm J}$\tablefootmark{d} & $\langle \lambda_{\rm sep}\rangle / \lambda_{\rm J}$\tablefootmark{e} & $M_{\rm J}$\tablefootmark{f} & $N_{\rm J}$\tablefootmark{g} \\  
       & [M$_{\sun}$] & & [pc] & [pc] & & [M$_{\sun}$] &\\ 
\hline
SMM 1 & $128\pm68$ & 2 & $0.19\pm0.01$ & $0.20\pm0.05$ & $0.95\pm0.24$ & $8.3\pm0.6$ & $15\pm8$\\
SMM 4 & $136\pm73$ & 3 & $0.23\pm0.04$ & $0.12\pm0.03$ & $1.92\pm0.58$ & $6.3\pm1.7$ & $22\pm13$\\
IRAS 13037 & $59\pm31$ & 2 & 0.41 & $0.16\pm0.04$ & $2.56\pm0.64$ & $12.3\pm3.2$ & $5\pm3$\\
SMM 6 & $34\pm18$ & 2 & $0.17\pm0.01$ & $<0.21$ & $>0.76$ & $<11.1$ & $>1$\\
SMM 7 & $41\pm22$ & 2\tablefootmark{h} & 0.09 & $0.15\pm0.04$ & $0.60\pm0.16$ & $8.3\pm2.2$ & $5\pm3$\\
IRAS 13039 & $58\pm31$ & 2 & $0.24\pm0.07$ & $0.22\pm0.06$ & $1.09\pm0.44$ & $18.5\pm4.8$ & $3\pm2$\\
\hline 
\end{tabular} 
\tablefoot{\tablefoottext{a}{Mass of the clump (see Table~\ref{table:laboca_properties}).}\tablefoottext{b}{Number of SABOCA cores in the LABOCA clump.}\tablefoottext{c}{The mean value of the nearest-neighbour distances between the cores. The quoted uncertainty is the standard error of the mean, and it could be calculated for systems of three or more cores.}\tablefoottext{d}{The local thermal Jeans length calculated for the clump temperature from Table~\ref{table:laboca_properties}.}\tablefoottext{e}{The ratio between the observed average core separation and the Jeans length.}\tablefoottext{f}{The Jeans mass corresponding to the Jeans length in column~(5).}\tablefoottext{g}{The Jeans number defined by $N_{\rm J}=M_{\rm J}/M$.}\tablefoottext{h}{A visual inspection of the SABOCA image suggests that SMM~7 is fragmented into two cores, but our {\tt BLOBCAT} analysis treated the cores as a single source.}  }
\end{minipage}  }
\end{table*}

\subsection{Star formation in G304.74+01.32}

Out of the 17 cores detected in the SABOCA map, plus the three LABOCA clumps that were resolved out in the SABOCA map (SMM 5, SMM 8, and BLOB 1), 13 appear to be associated with a YSO (or YSOs) on the basis of the \textit{WISE} data (two sources exhibit \textit{WISE} shock emission). Hence, the percentage of star-forming cores in G304.74 is $65\% \pm 18\%$. The remaining seven sources ($35\% \pm 13\%$) appear dark in the \textit{WISE} IR images, and can be considered candidate starless cores. However, owing to the distance to G304.74, $d=2.54\pm0.66$~kpc, it is possible that the apparently IR-dark sources contain deeply embedded, low-luminosity YSOs, which remain below the detection limit of the \textit{WISE} data. On the other hand, the dust optical thickness in the \textit{WISE} W4 band at 22~$\mu$m can be high enough to render the embedded IR sources undetected (e.g. \cite{ragan2012}). For comparison, Henning et al. (2010) found that 11 out of their 18 cores ($61\% \pm 18\%$) in the Snake IRDC are associated with \textit{Spitzer} 24~$\mu$m emission, which is suggestive of a similar percentage of star-forming cores as in G304.74. Ohashi et al. (2016) found that 20/48, or $42\% \pm 9\%$ of the cores identified with the Atacama Large Millimetre/Submillimetre Array (ALMA) at 3~mm in the IRDC G14.225-0.506 are star-forming, which is also consistent with our result within the Poisson counting uncertainties. Similarly, Zhang et al. (2017) found that 18/44 of their sample of IRDC cores show evidence of molecular outflows, which suggests a star-forming percentage of $41\% \pm 10\%$ among the cores.

\begin{figure}[!htb]
\centering
\resizebox{0.98\hsize}{!}{\includegraphics{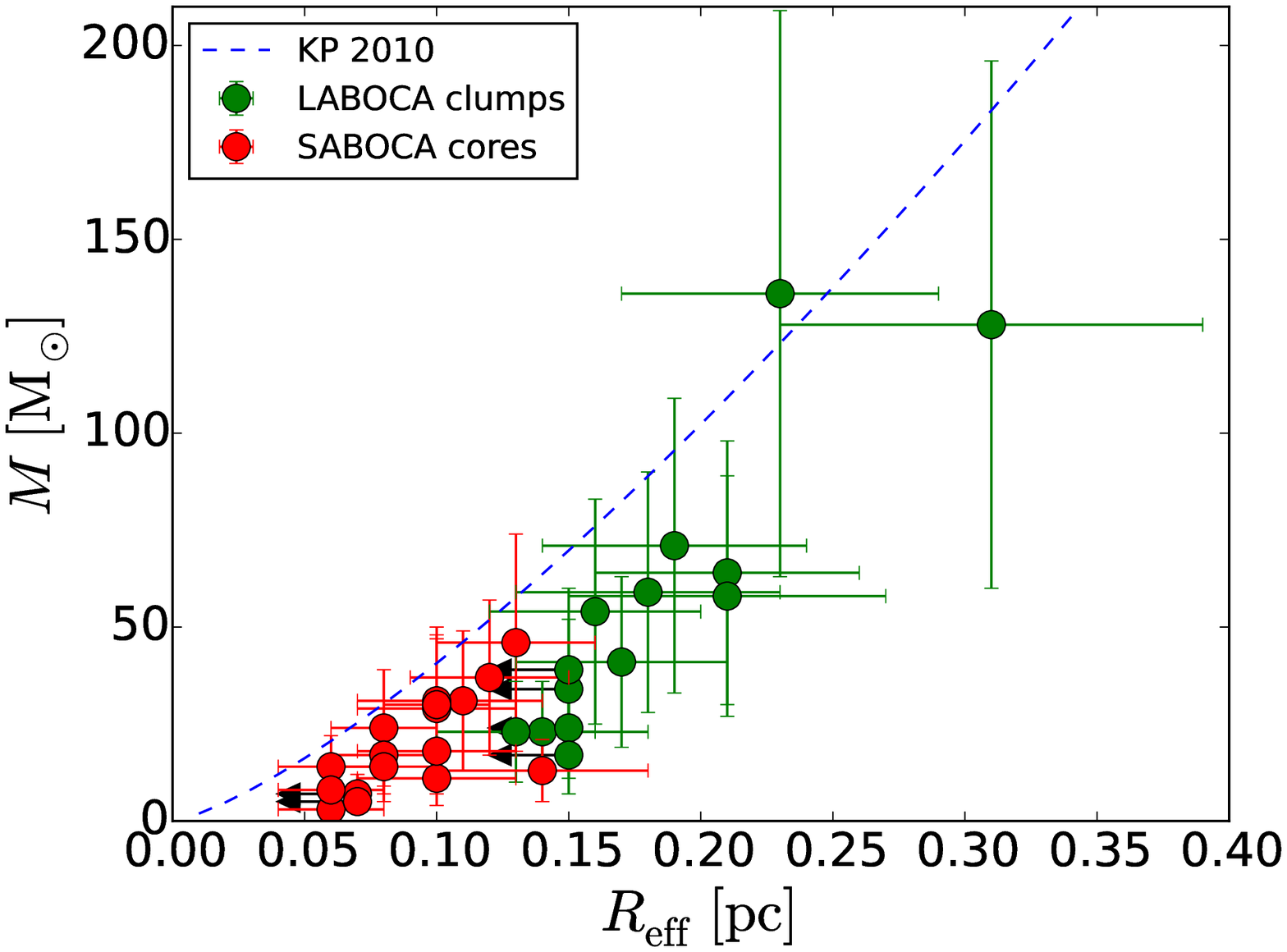}}
\caption{Masses of the LABOCA clumps and SABOCA cores in G304.74 plotted as a function of their effective radii. The upper size limits are indicated by left pointing arrows. The blue dashed curve represents the mass-radius threshold for high-mass star formation proposed by KP10, that is $M(R)=870\, {\rm M}_{\sun} \times (R/{\rm pc})^{1.33}$.}
\label{figure:massradius}
\end{figure}

In Fig.~\ref{figure:massradius}, we plot the masses of our LABOCA clumps and SABOCA cores as a function of their effective radii. These values are compared with the threshold for high-mass star formation derived by KP10, which is indicated in Fig.~\ref{figure:massradius}. Considering the nominal values of $M$ and $R_{\rm eff}$ of our LABOCA clumps, only one of them, SMM~4, appears to lie above the aforementioned threshold, and hence capable of forming a massive star(s). On the other hand, none of the SABOCA cores appear to fulfil the KP10 threshold. However, owing to the large uncertainties in the derived masses and sizes of the clumps and cores, some of them could actually lie above the KP10 curve in Fig.~\ref{figure:massradius}. We note, however, that the mass uncertainty was partly propagated from the distance uncenrtainty, and while a larger distance would imply a higher mass ($M \propto d^2$), the physical size of the source would also be larger ($R_{\rm eff} \propto d$), which would move a data point to the right in Fig.~\ref{figure:massradius}. Hence, the error bars in Fig.~\ref{figure:massradius} are not independent of each other. 

Using radio continuum observations at 18~GHz and 22.8~GHz, S\'anchez-Monge et al. (2013) found that there is an optically thin \ion{H}{ii} region associated 
with IRAS~13039 in G304.74. The radio peak position lies $14\farcs5$ to the south-east of our SABOCA core IRAS~13039a (see Fig.~\ref{figure:fragments}). The authors did not detect 22.2~GHz water maser line emission towards the \ion{H}{ii} region, and the source was classified as an early-stage Type~1 object (a rare case among their source sample, because only $8\%$ of Type~1 objects were found to be associated with \ion{H}{ii} regions). The spectral type of the central ionising star was concluded to be B1 (they adopted a distance of 2.4~kpc for IRAS~13039). Hence, the IRAS~13039 region appears to be associated with high-mass star formation, although the corresponding LABOCA clump or the SABOCA core IRAS~13039a do not fulfil the KP10 threshold (not even within the error bars plotted in Fig.~\ref{figure:massradius}). The IRAS~13039 region is surrounded by an extended, diffuse arc-like emission feature in the \textit{WISE} 12~$\mu$m and 22~$\mu$m images (Fig.~\ref{figure:wise}). This could be an indication of a photodissociation region (PDR) surrounding the associated \ion{H}{ii} region, that is a boundary between the fully ionised region and the surrounding neutral gas. The PDRs are expected to be extended and strong emitters at 12~$\mu$m. Alternatively (or additionally), 
the observed features in the \textit{WISE} 12~$\mu$m and 22~$\mu$m images 
could be related to the shocks driven by stellar winds (\cite{peri2012}). Hence, the star formation activity in IRAS~13039 can have some feedback effect on the nearby clumps (e.g. SMM~8) that can be either negative or positive in terms of star formation, that is it can either prevent or slow down the clump collapse to YSOs, or trigger star formation.

Apart from IRAS~13039, it is unclear whether the other clumps or cores in G304.74 are forming high-mass stars, or low to intermediate-mass stars. Construction and analysis of the SEDs of the detected SABOCA sources associated with the \textit{WISE} detected YSOs would help to answer this question. For example, although the SMM~4 clump lies above the KP10 threshold for high-mass star formation, the clump is strongly fragmented (Fig.~\ref{figure:fragments}), which might prefer the formation of a cluster of lower mass stars. However, competitive accretion in such a system could eventually lead to the formation of a high-mass star (\cite{bonnell2001}; \cite{bonnell2006}). Also, the aforementioned case of IRAS~13039 demonstrates that clump fragmentation does not necessarily prevent the formation of massive stars. Finally, if the clump or core can accrete more mass from the surrounding medium, for example via the filamentary striations seen in the \textit{Herschel}/SPIRE images towards G304.74 (Fig.~\ref{figure:spire}), the central YSO embedded in the source can grow to become a high-mass star (e.g. \cite{wang2011}; \cite{ohashi2016}).  

\subsection{The Seahorse Nebula in a wider context of infrared dark clouds}

The studies presented in Papers~I and II, together with the present results, allow us to discuss the IRDC G304.74 in a broader context of Galactic IRDCs and their nature as stellar nurseries. One fairly atypical property of G304.74 is its spatial location in the Galaxy. The Galactic latitute of G304.74, $b=+1 \fdg 32$, places it about 59~pc above the Galactic plane at the near kinematic distance (Paper~II, Appendix~A.1 therein). This approaches the scale height of the molecular disk, that is $z_{1/2} \sim 70$~pc (Bronfman et al. 1988). On the other hand, the Galactocentric distance of G304.74 is about 7.3~kpc. For comparison, Jackson et al. (2008) found a peak in the fourth quadrant IRDC number density at a Galactocentric distance of $\sim6$~kpc, while Finn et al. (2013) found that the first quadrant IRDCs are concentrated at a Galactocentric distance of $\sim4.5$~kpc (see also \cite{simon2006b}). These results are expected to be an indication that the IRDCs are associated with a Galactic spiral arm structure. The Galactocentric distance of the fourth Galactic quadrant IRDC G304.74 is over $20\%$ higher than that of typical IRDCs found by Jackson et al. (2008), which is consistent with the finding that G304.74 could not be linked to a spiral arm (Sect.~1.).

The IRDC G304.74 is filamentary in shape, which is a fairly common characteristic among IRDCs. In particular, G304.74 appears remarkably similar to the IRDC G11.11-0.12, or the Snake Nebula, although the Snake is over three times longer in projection ($\sim30$~pc; e.g. \cite{kainulainen2013}). For comparison, the IRDC G338.4-0.4, or the Nessie Nebula, has an extreme projected length of $\sim80$~pc with an aspect ratio of $r_{\rm aspect} \sim150:1$ (\cite{jackson2010}). Jackson et al. (2010) used the HNC$(1-0)$ rotational line emission to estimate the line mass of the Nessie, and obtained a value of 110~M$_{\sun}$~pc$^{-1}$ under the assumption that the HNC abundance is $2\times10^{-9}$. Although the Nessie's line mass estimate is based on a different method than we have used, and suffers from the uncertain HNC abundance, it is comparable to that of the G304.74 LABOCA filament ($M_{\rm line}^{\rm LABOCA}=87\pm27$~M$_{\sun}$~pc$^{-1}$). In contrast, Kainulainen et al. (2013) derived a high line mass of $\sim600$~M$_{\sun}$~pc$^{-1}$ for the Snake IRDC, where the cloud mass was derived from near and mid-IR extinction data. Because the extinction technique is sensitive to low-column density material missed by bolometer dust continuum maps, a direct comparison with the present results cannot be done.

That many of the IRDCs are filamentary in shape suggests that they might have formed through ubiquitous, converging supersonic turbulent flows as seen in numerical simulation (e.g. \cite{klessen2001}; \cite{banerjee2006}; \cite{girichidis2012}; \cite{federrath2013}; \cite{smith2014}, and references therein). Although the detailed formation mechanism(s) of self-gravitating filamentary structures remains to be understood, the observed striations around filamentary clouds, such as those seen in the present study, provide valuable insights into their formation and evolutionary processes (\cite{palmeirim2013}; \cite{cox2016}; \cite{saajasto2017}, and references therein). In particular, the main filament can increase its mass through accretion along the striations. As in the case of the G304.74 system, the striations are typically found to be oriented perpendicular to the main filament. Moreover, the striations are found to be aligned along the magnetic field direction, and hence the main filament and the associated magnetic field are supposed to be perpendicular to each other as well (\cite{palmeirim2013}; \cite{chen2014}; \cite{inutsuka2015}). If this is indeed the case, the cloud might have originated in a self-gravitating gas layer where the magnetic field lies in the plane of the layer, and which then fragments into thermally supercritical filaments threaded by perpendicular magnetic field lines (see \cite{nagai1998}; \cite{cox2016}). The perpendicular striations could then represent the remnants of the original gas layer that can feed the growth of the newly formed filament. In principle, the comparatively low LABOCA line mass of the Seahorse Nebula suggests that it might be in an early stage of the accretion process, although the true mass of the filament is expected to be higher owing to the low-column density material missed by our observations.  

The observational studies of the substructure characteristics of filamentary IRDCs, such as the Nessie (\cite{jackson2010}), the Snake (\cite{kainulainen2013}), and G11.36+0.80 (\cite{miettinen2012b}), strongly support the scenario where the fragmention of the parent filament into clump-sized units is driven by cylindrical fragmention, most notably via sausage fluid instability. Our results suggest that the Seahorse Nebula G304.74 is no exception: its clumps are projectively separated in a fashion predicted by the sausage instability paradigm.

As the present study of the Seahorse Nebula demonstrates, IRDCs exhibit multi-scale fragmentation. On the spatial scale probed by our SABOCA observations, the fragmentation of the clumps into smaller cores is consistent with a process of gravitational Jeans instability within the measurement uncertainties. On the other hand, in terms of angular resolution, the present study links the more typical, single-dish bolometer studies of poorer resolution to the higher resolution interferometric studies. The latter types of studies have revealed still smaller objects within IRDC clumps, and hence the hierarchical, multi-scale nature of the fragmentation of IRDCs (e.g. \cite{wang2014}; \cite{ragan2015}; \cite{zhang2015}; \cite{beuther2015}; \cite{busquet2016}; \cite{henshaw2016}; \cite{ohashi2016}; \cite{kong2017a},b; \cite{henshaw2017}; \cite{sanhueza2017}). For example, the 1.07~mm, $1\farcs4 \times 0\farcs8$ resolution ALMA observations by Henshaw et al. (2017) revealed cores with a mean effective radius of $\sim0.02$~pc in their target clump, while Sanhueza et al. (2017), who used 1.3~mm, $3\farcs5$ resolution Submillimetre Array observations, resolved their target clump into five fragments with radii in the range of 0.019--0.043~pc. Sanhueza et al. (2017) concluded that the clump fragmentation is not predominantly driven by thermal or turbulent pressures. Indeed, the small-scale fragmentation of IRDCs might involve properties that cannot be captured by simple, Jeans-type gravitational fragmentation models, such as complex density profiles and the interplay between turbulence and magnetic fields (e.g. \cite{beuther2015}). Nevertheless, high-resolution observations of IRDCs are required to improve our still incomplete understanding of the fragmentation of IRDCs down to $\sim0.01$~pc scales.

Apart from the IRAS~13039 \ion{H}{ii} region in G304.74, the currently available observational data are not sufficient to tell whether the cloud could give birth to high-mass stars. However, the evidence of ongoing high-mass star formation in many of the Galactic IRDCs is very persuasive. This includes the presence of Class~II methanol and water maser emission (\cite{pillai2006}; \cite{wang2006}; \cite{chambers2009}), hot molecular cores (e.g. \cite{rathborne2008}), embedded \textit{Spitzer} 24~$\mu$m sources (e.g. \cite{chambers2009}), and ultracompact \ion{H}{ii} regions (e.g. \cite{battersby2010}). The presence of high-mass star formation in IRDCs is also consistent with the finding that the clouds are typically located in or near the Galactic spiral arms (\cite{finn2013}). As discussed in Sect.~4.3, G304.74, which does not appear to be associated with a spiral arm, might represent the formation site of mostly low and intermediate-mass stars. Alternatively, at least some of the clumps and cores along the filament (besides IRAS~13039) might increase their mass sufficiently much to allow future high-mass star formation. Hence, G304.74 is a promising target source to investigate the physical conditions of clump-fed, competitive accretion mode of star formation.

\section{Summary and conclusions}

We used the SABOCA bolometer to map the filamentary IRDC~G304.74+01.32 in the 350~$\mu$m dust continuum emission. These data were used in conjunction with our previous LABOCA 870~$\mu$m data for the cloud, and the \textit{Herschel} far-IR to submm and \textit{WISE} near-IR to mid-IR imaging data. Our main results are summarised as follows: 

\begin{enumerate}
\item The SABOCA imaging at $9\arcsec$ resolution revealed that $36\%\pm16\%$ of the LABOCA clumps in G304.74 are fragmented into two or more cores, but the morphology of some of the identified SABOCA cores suggests that this multiplicity fraction could be somewhat higher, namely $43\%\pm17\%$. On a larger scale, our SABOCA imaging revealed a dense, narrow 
($0.18\pm0.05$~pc on average) central filament inside the LABOCA filament. 
\item On the basis of the \textit{WISE} IR imaging data, $65\%\pm18\%$ of the SABOCA cores are hosting YSOs, while the remaining $35\%\pm13\%$ appear dark in the \textit{WISE} IR images. The latter IR-dark cores are candidates for prestellar cores. 
\item The mean dust temperature of the clumps, as derived using the \textit{Herschel}/SPIRE 250, 350, and 500~$\mu$m data, was found to be $15.0 \pm 0.8$~K. The mean mass, beam-averaged H$_2$ column density, and H$_2$ number density of the LABOCA clumps are estimated to be $55\pm10$~M$_{\sun}$, $(2.0\pm0.2)\times10^{22}$~cm$^{-2}$, and $(3.1\pm0.2)\times10^4$~cm$^{-3}$. The corresponding values for the SABOCA cores are $29\pm3$~M$_{\sun}$, $(2.9\pm0.3)\times10^{22}$~cm$^{-2}$, and $(7.9\pm1.2)\times10^4$~cm$^{-3}$. Hence, on average, the cores are found to be about 1.9 times less massive, and to have a factor of 1.5 (2.5) times higher column (number) density than their parent LABOCA clumps.
\item The filamentary structure of G304.74 is estimated to be thermally supercritical by a factor of at least 3.5 for the LABOCA filament, and by a factor of $ \gtrsim1.5$ for the SABOCA filament. If non-thermal motions, or turbulence, are important on a larger scale traced by LABOCA, the corresponding filamentary structure could be closer to virial equilibrium. Nevertheless, the presence of substructure and YSOs along the whole long axis of the cloud shows that it has fragmented owing to gravitational instability, which further argues for G304.74 being thermally supercritical. 
\item The \textit{Herschel}/SPIRE images showed that G304.74 is associated with elongated striations that connect to the filament. This could be a sign that G304.74 is still accreting mass from the surrounding medium. Hence, the pressure of the ambient medium might play an imporant role in the dynamical evolution of the G304.74 filament. 
\item Our fragmentation analysis suggests that the LABOCA clumps might have fragmented into cores via thermal Jeans-type instability, although contribution from non-thermal pressure support cannot be excluded on the basis of the present data. Nevertheless, the hierarchical fragmentation of G304.74 might have proceeded from the formation of clumps through sausage instability, to a gravitational Jeans instability on subclump scales. 
\item One of the clumps in G304.74, IRAS 13039-6108 in the centre of the filament, is known to be associated with an \ion{H}{ii} region, and hence the clump is hosting a high-mass YSO(s). Further studies are needed to understand if some of the other clumps and cores in G304.74 have the potential to give birth to massive stars, or whether they are more likely to represent the formation sites of low and intermiediate-mass stars. Because G304.74 is a relatively nearby IRDC ($d \sim 2.5$~kpc), it could be used as a useful target source to investigate the formation of intermediate-mass stars, and how the process compares with those of low and high-mass star formation. Owing to the apparent morphology of G304.74 in the SPIRE maps, we propose that the cloud is nicknamed the Seahorse Nebula.
\end{enumerate}

\begin{acknowledgements}

I thank the referee for providing insightful and constructive comments that helped to improve 
the quality of this paper. I am grateful to the staff at the APEX telescope for performing the 
service mode SABOCA and LABOCA observations presented in this paper. 
The research presented in this article was funded by the Academy of Finland under project 
no.~132291, and the European Union's Seventh Framework 
Programme (FP7/2007--2013) under grant agreement 337595 (ERC Starting Grant, 'CoSMass'). 
This research has made use of data from the \textit{Herschel} Gould Belt survey (HGBS) 
project (\url{http://gouldbelt-herschel.cea.fr}). The HGBS is a \textit{Herschel} Key 
Programme jointly carried out by SPIRE Specialist Astronomy Group 3 (SAG 3), scientists 
of several institutes in the PACS Consortium (CEA Saclay, INAF-IFSI Rome and 
INAF-Arcetri, KU Leuven, MPIA Heidelberg), and scientists of the \textit{Herschel} 
Science Center (HSC). This publication makes use of data products from 
the \textit{Wide-field Infrared Survey Explorer}, which is a joint project of 
the University of California, Los Angeles, and the Jet Propulsion Laboratory/California 
Institute of Technology, funded by the National Aeronautics and Space Administration. 
This research made use of Astropy, a community-developed core Python package for Astronomy 
(\cite{astropy2013}). This research has made use of NASA's Astrophysics 
Data System and the NASA/IPAC Infrared Science Archive, which is operated by 
the JPL, California Institute of Technology, under contract with the NASA.

\end{acknowledgements}

\appendix

\end{document}